\documentclass[aip,jcp,amsmath,amssymb,reprint,numeric]{revtex4-1}
\usepackage{graphicx}
\usepackage{subfigure}
\usepackage{color}
\usepackage{amsmath}
\usepackage{bm}

\begin{document}
\title {Effect of  total and pair configurational entropy in determining dynamics of supercooled liquids over a range of densities}
\author{Atreyee Banerjee}

\affiliation{\textit{Polymer Science and Engineering Division, CSIR-National Chemical Laboratory, Pune-411008, India}}

\author{Manoj Kumar Nandi}
\affiliation{\textit{Polymer Science and Engineering Division, CSIR-National Chemical Laboratory, Pune-411008, India}}

\author{ Srikanth Sastry}
\affiliation{\textit{Theoretical Sciences Unit, Jawaharlal Nehru Centre for Advanced Scientific Research, Jakkur Campus, Bengaluru 560 064, India}}
\author{Sarika Maitra Bhattacharyya}
\email{mb.sarika@ncl.res.in}
\affiliation{\textit{Polymer Science and Engineering Division, CSIR-National Chemical Laboratory, Pune-411008, India}}

\date{\today}

\begin{abstract} 
In this paper,  we present a study of supercooled liquids interacting with the Lennard Jones (LJ) potential and the 
corresponding purely repulsive  (Weeks-Chandler-Andersen or WCA) potential,
over a range of densities and temperatures, in order to
 understand the origin of their different dynamics
 in spite of their structures being similar. Using the configurational entropy as the thermodynamic marker  via the Adam Gibbs (AG) relation, 
we show that the difference in the dynamics of
 these two systems at low temperatures can be explained from thermodynamics. At higher densities both the thermodynamical and dynamical difference 
between these model systems decrease, which is quantitatively demonstrated
 in this paper by calculating different parameters. 
 The study also reveals the origin of the difference in pair entropy despite the similarity in the structure.
Although the maximum difference in structure is obtained in the partial radial distribution function (rdf) of the B type of particles, 
the rdf of AA pairs and AB pairs give rise to the 
differences in the entropy and dynamics.
This work supports the observation made in an earlier study (Phys. Rev. Lett.,\textbf {113}, 225701, 2014) and shows that they are generic in nature,
 independent of density.

\end{abstract}
\maketitle


\section{Introduction}

Historically, the van der Waals picture assumes that the short range, harsh repulsive part of the 
potential determines the structure of the fluid and
 the attractive part of the potential provides a homogeneous background \cite{hansen-mac}. 
In this spirit, Weeks, Chandler and Andersen proposed that for liquids at high density, the Lennard 
Jones (LJ) potential can be treated as a sum of repulsive and attractive parts, and the repulsive part of the
 potential, commonly known as the WCA potential, is sufficient to describe the 
structure and the dynamics of liquids \cite{chandler}. 
 For a supercooled liquid with closely packed particles this picture is expected to  provide a good description.
 However, in a series of papers Bertheir and  Tarjus have shown that 
although the structures of the WCA and LJ systems are very close at the same temperature and density, their dynamics at low temperatures are very different 
\cite{tarjus_prl,tarjus_pre,tarjus_epje,tarjus_berthier_jcp}. 
They have questioned the role of thermodynamics in determining the dynamics. They have also used various theories which depend on the pair correlation
function and have shown that none of the theories can differentiate the two systems\cite{tarjus_epje, tarjus_pre,Schweizer1,Schweizer2}.
Thus they have proposed that higher-order correlation functions or topological
measure are more sensitive to capture the difference between the two systems\cite{tarjus_berthier_jcp}.
Coslovich has indeed shown that triplet correlations are much more sensitive
and can disentangle the two systems \cite{coslovich,coslovich-jcp}.  He has shown that the LJ system
 has more pronounced 
local ordering \cite{coslovich}. The sluggish dynamics in supercooled liquids comes from these locally preferred structures 
which are known to form correlated domains \cite{tarjus-kivelson}. 
The estimation of the lengthscale of the domains is done in different ways \cite{smarajit_srikanth_pnas,Smarajit_pre,saroj_sriram_prl,Paddy-jcp13, hocky-prl,kim_saito}. 
One of them is the measurement of the point-to-set (PTS) correlation length \cite{RFOT_cavagna,RFOT_cavagna2,Chiara_PNAS_2012,hocky-prl}.
Hockey {\it et. al} have shown that the difference in the static length scale (PTS) can account for the difference in the dynamics of the LJ and WCA systems.
Although the connection between the  PTS and dynamics does point to the role of thermodynamics, it cannot completely rule out the  role of kinetic constraints\cite{Jack_garrahan}. 
Wang et al. have connected the glass transition temperature to the vibrational density of states and they found that above a critical density,
the LJ and WCA systems behave in a similar way\cite{Wang-Xu-prl}.
A similar connection between the two systems have also been made by looking at fragility\cite{Wang-soft-matter}.
Dyre and co-workers \cite{Toxvaerd, Pedersen-prl,Bohling} 
have argued that the origin of dynamical differences between LJ and WCA systems is not the neglect of the attractive part \cite{Bohling} but rather  
the truncation of the WCA potential\cite{Toxvaerd}.
They have also predicted that at high densities when there is inclusion of interactions from all first neighbours, the WCA system can mimic the LJ system. 
Bohling {\it et al.}  have also shown that the dynamics of the attractive LJ system can be mapped to a repulsive inverse power law potential proving their point that the
difference in the dynamics is not due to the nature of the potential \cite{Bohling}.

In a recent work by some of us, it has been shown that the dynamics of the LJ and WCA systems follow the thermodynamics. It has been conclusively proven that
the difference in the dynamics is  a reflection of the difference in their thermodynamics
\cite{bssb}. The well known Adam-Gibbs relation\cite{adam-gibbs}, which connects thermodynamics to kinetics 
\cite{sciortino-kob-tartaglia-prl,Srikanth_nature,speedy-1,speedy-2,foffi,stanley-nature,starr-sciortino-et-al,saika-voivod,mossa-tartaglia-pre,starr_douglas_sastry},
is found to be valid for LJ and WCA systems. It has also been found that although 
their pair correlation functions look
very similar\cite{tarjus_prl}, 
the pair entropy and pair relaxation time predicted by  Adam-Gibbs relaxation are capable of showing the difference between the LJ and WCA systems\cite{bssb}.
However  our previous study was performed at a specific density.
 In the present paper we have studied a wider range of 
densities, and show that the observations made at $\rho = 1.2$ are generic.
 We study the density dependence of different thermodynamic and kinetic parameters such as the 
fragility \cite{Angell_nature,Ito_Angell} and various transition temperatures. 
Here we analyse the origin of the difference in the pair entropy between the two models.
 We also show that although the AG relation at the two body level shows reminiscence of mode 
coupling theory (MCT), it fares better in predicting the difference
between the dynamics of the two systems.

The paper is organized as follows:
The simulation details are given in Sec. II. In Sec. III we describe
the methods used for evaluating the various quantities of interest and provide other necessary background. In Sec. IV we report the simulation results and 
their analysis. Sec. V contains a discussion of
 presented results and conclusions. 
\section{Simulation Details}
We have performed molecular dynamics simulations of the Kob-Andersen model which is a binary mixture (80:20) of Lennard-Jones (LJ) particles and the 
corresponding WCA version \cite{kob,chandler}. 
 The interatomic pair  
potential between species $\alpha$ and $\beta$, with ${ \alpha,\beta}= A,B$, 

$U_{\alpha\beta}(r)$ is described by a  truncated and shifted Lennard-Jones (LJ) potential, as given by:
\begin{widetext}
\begin{equation}
 U_{\alpha\beta}(r)=
\begin{cases}
 U_{\alpha\beta}^{(LJ)}(r;\sigma_{\alpha\beta},\epsilon_{\alpha\beta})- U_{\alpha\beta}^{(LJ)}(r^{(c)}_{\alpha\beta};\sigma_{\alpha\beta},\epsilon_{\alpha\beta}),    & r\leq r^{(c)}_{\alpha\beta}\\
   0,                                                                                       & r> r^{(c)}_{\alpha\beta}
\end{cases}
\end{equation}
\end{widetext}

\noindent where $U_{\alpha\beta}^{(LJ)}(r;\sigma_{\alpha\beta},\epsilon_{\alpha\beta})=4\epsilon_{\alpha\beta}[({\sigma_{\alpha\beta}}/{r})^{12}-({\sigma_{\alpha\beta}}/{r})^{6}]$ and
 $r^{(c)}_{\alpha\beta}=2.5\sigma_{\alpha\beta}$ for the LJ systems and $r^{(c)}_{\alpha\beta}$  is equal to the position of the minimum of $U_{\alpha\beta}^{(LJ)}$
for the WCA systems. Length, temperature and
time are given in units of $\sigma_{AA}$, ${k_{B}T}/{\epsilon_{AA}}$ and $\tau = \surd({m_A\sigma_{AA}^2}/{\epsilon_{AA}})$, 
respectively.  
Here we have simulated the Kob-Andersen Model  
with the interaction parameters  $\sigma_{AA}$ = 1.0, $\sigma_{AB}$ =0.8 ,$\sigma_{BB}$ =0.88,  $\epsilon_{AA}$ =1, $\epsilon_{AB}$ =1.5,
 $\epsilon_{BB}$ =0.5, $m_{A}$ = $m_B$=1.0 .

The molecular dynamics (MD) simulations have been carried out using the LAMMPS 
package \cite{lammps}.
We have performed MD simulations in the canonical ensemble (NVT) using  Nos\'{e}-Hoover thermostat  with integration timestep 0.005$\tau$. The time
constant for the Nos\'{e}-Hoover thermostat  are taken to be 100  timesteps.
The sample is kept in a cubic box with periodic boundary conditions.
 The system size is $N = 500$, $N_A = 400$ (N $=$ total number
of particles, $N_A$ $=$ number of particles of type A) and we have studied a broad range of density $\rho$ from 1.2 to
1.6. For all state points, three to five independent samples with run lengths $>$ 100$\tau$ ($\tau$ is the $\alpha$-
relaxation time) are analyzed.

\section{Definition}
\subsection{Relaxation time}
We have calculated the relaxation times from the decay of the
overlap function q(t), from the condition $q(t = \tau_{\alpha} , T )/N =
1/e$.  $q(t)$ is defined as
\begin{eqnarray}
\langle q(t) \rangle \equiv \left \langle \int dr \rho(r, t_0 )\rho(r, t + t_0 )\right \rangle \nonumber\\
=\left \langle \sum_{i=1}^{N}\sum_{j=1}^{N} \delta({\bf{r}}_j(t_0)-{\bf{r}}_i(t+t_0)) \right \rangle \nonumber\\
=\left \langle \sum_{i=1}^{N} \delta({\bf{r}}_i(t_0)-{\bf{r}}_i(t+t_0)) \right \rangle \nonumber\\
+\left \langle \sum_{i}\sum_{j\neq i} \delta({\bf{r}}_i(t_0)-{\bf{r}}_j(t+t_0)) \right \rangle.
\end{eqnarray}
The overlap function is a two-point time correlation
function of local density $\rho(r, t)$. It has been used in
many recent studies of slow relaxation \cite{shila-jcp}.
 In this
work, we consider only the self-part of the total overlap
function (i.e. neglecting the $i \neq j$ terms in the double
summation). Earlier it has been shown to be a good approximation to the full
 overlap function. So,
the overlap function can be well approximated by its self part, 
and written as,  
\begin{eqnarray}
 \langle q(t) \rangle \approx \left \langle \sum_{i=1}^{N} \delta({\bf{r}}_i(t_0)-{\bf{r}}_i(t+t_0)) \right \rangle.
\end{eqnarray}

Again, the $\delta$ function is approximated by a window function $\omega(x)$ which defines the
condition of “overlap” between two particle positions
separated by a time interval t:
\begin{eqnarray}
 \langle q(t) \rangle \approx \left \langle \sum_{i=1}^{N} \omega (\mid{\bf{r}}_i(t_0)-{\bf{r}}_i(t+t_0)\mid) \right \rangle \nonumber\\
\omega(x) = 1, x \leq {\text{a implying “overlap”}} \nonumber\\
=0, \text{otherwise}.
\end{eqnarray}

The time dependent overlap function thus depends on
the choice of the cut-off parameter a, which we choose
to be 0.3. This parameter is chosen such that particle positions separated due to small amplitude vibrational motion are treated as the same, or that $a^2$ is comparable to
the value of the MSD in the plateau between the ballistic
and diffusive regimes.

  Relaxation times obtained from the decay of the self intermediate scattering function $F_s (k, t)$ using the 
definition $F_s (k, t = \tau_\alpha , T)$ = $1/e$ at $k\simeq 2\pi /r_{min}$ . The self
intermediate scattering function is calculated from the
simulated trajectory as
\begin{equation}
 F_s(k,t)=\frac{1}{N}\left \langle \sum_{i=1}^{N} \exp(-i{\bf{k}}.({\bf{r}}_i(t)-{\bf{r}}_i(0))) \right \rangle.
\end{equation}

Since relaxation times from $q(t)$ and $F_s (k, t)$
behave very similarly at low temperature, we have used the time scale
obtained from q(t). 

\subsection{Diffusivity}
Diffusivities (D ) are obtained from the mean squared
displacement (MSD) of the particles. We have calculated MSD as follows,
\begin{equation}
 R^2(t)=\frac{1}{N} \sum_{i} \langle({\bf{r}}_i(t)-{\bf{r}}_i(0))^2 \rangle, 
\end{equation}
 and from the long time behavior of MSD, the diffusion coefficient D obtained as,
\begin{equation}
 D= \lim_{t\rightarrow \infty} \frac{R^2(t)}{6t}.
\end{equation}

\subsection{Fragility}
Fragility  \cite{Angell_nature,Ito_Angell} is a parameter which quantifies the rate at which any  dynamical quantity (viscosity, relaxation
times and inverse diffusivities) grows  with temperature.
There are many ways to quantify the fragility among which two are more popular. One is from 
 the ``steepness index'' m, which is defined from the so-called Angell plot. The slope (m), of logarithm of the viscosity ($\eta$)
or relaxation time ($\tau$ ) at the laboratory glass transition
temperature $T = T_g$ , with respect to the scaled inverse
temperature $\frac{T_g}{T}$, is written as, $ m = (\frac{ d \log \tau }{d(T_g/T)})_{T=T_g}$

Another way  of defining fragility is by  plotting the temperature dependence of the viscosity or the relaxation time  in terms of modified Vogel-Fulcher-Tammann (VFT) 
equation, 
\begin{equation}
\tau(T)=\tau_{o}\exp \left[ \frac{1}{K_{VFT}(\frac{T}{T_{VFT}}-1)}\right],
\label{VFT}
\end{equation}
\noindent
where $K_{VFT}$ is the kinetic marker for fragility, $T_{VFT}$ is the temperature where the relaxation time diverges and $\tau_{o}$ is the high temperature relaxation time.
In this paper we have used the Vogel-Fulcher-Tammann (VFT) fits to the temperature dependence of dynamical quantities for defining kinetic fragility.

The thermodynamic fragility, on the other hand, is obtained from the temperature dependence of the configurational entropy $S_{c}$. The method of $S_c$ 
calculation will be discussed in the next subsection.
The temperature
dependence of $S_c$ is given by
\begin{equation}
TS_{c}=K_{T} \left(\frac{T}{T_{K}}-1\right),
\label{TS}
\end{equation}
\noindent 
  where $K_{T}$ is termed as the thermodynamic marker for fragility  and $T_K$ is the Kauzmann temperature.
 The correlation between the kinetic fragility and the thermodynamic fragility and thus the potential energy landscape can be drawn from the well known Adam-Gibbs 
(AG) relation \cite{adam-gibbs},
\begin{equation}
\tau(T)=\tau_{o}\exp\left(\frac{A}{TS_{c}}\right).
\label{ag}
\end{equation}
\noindent
Here $A$ is the Adam-Gibbs parameter.  The above equation predicts Arrhenius behaviour for constant $S_{c}$ and thus it relates $A$  to the high temperature
 activation energy $E_{\infty}$ \cite{shila-jcp}. 
By using Eq.\ref{TS} in Eq.\ref{ag} the AG relation yields the VFT (Eq.\ref{VFT}) relation with the identification of
 $K_{VFT}\simeq K_{AG} = K_T/A$ and $T_{VFT}\simeq T_K$.  

\subsection{Configurational entropy}

Configurational entropy, $S_c$ per particle, the measure of
the number of distinct local energy minima, is calculated \cite{srikanth_PRL}
by subtracting from the total entropy of the system the
vibrational component:
$S_c (T ) = S_{total} (T ) - S_{vib} (T )$ \cite{shila-jcp,Srikanth_nature}.
Here the total entropy, $S_{total}=S_{id}+S_{ex}$, 
is obtained by summing the ideal gas entropy $S_{id}$ with the excess entropy $S_{ex}$ where the latter is  obtained via thermodynamic integration from the ideal gas limit.
 Vibrational entropy for the binary mixture  is calculated by making a harmonic approximation to the
 potential energy about a given local minimum.

\subsection{Pair configurational entropy}
As discussed earlier the excess entropy $S_{ex}$, defined as the difference between the total entropy ($S_{total}$) and the ideal gas entropy ($S_{id}$ ) at
the same temperature ($T$) and density ($\rho$), can also be expanded in an infinite series, 
$S_{ex}=S_{2}+S_{3}+.....=S_{2}+\Delta S$ using Kirkwood's factorization \cite{Kirkwood} of the N-particle distribution function \cite{green_jcp,raveche,Wallace}. 
 $S_{n}$ is the $``n"$ body contribution to the entropy. Thus the pair excess entropy is $S_{2}$ and the higher order contributions to excess entropy is given by 
the residual multiparticle entropy (RMPE), $\Delta S=S_{ex}-S_{2}$  \cite{giaquinta-1,giaquinta-2,giaquinta-3,giaquinta-4}.
The pair entropy $S_{2}$ for a binary system can be written in terms of 
the partial radial distribution functions,
\begin{widetext}
\begin{equation}
\frac{S_{2}}{k_{B}}=-2\pi\rho \sum_{\alpha,\beta}x_{\alpha} x_{\beta} \int_0^{\infty} \{g_{{\alpha}{ \beta}}(r) \ln g_{{\alpha} {\beta}}(r)- [g_{{\alpha}{\beta}}(r)-1]\} r^2 d {r},
\label{s2_final}
\end{equation}
\end{widetext}
\noindent  where $ g_{{\alpha}{ \beta}}(r)$ is the atom-atom pair correlation between atoms of type $\alpha$ and $\beta$, $\rho$ is the density of the system,
 $x_{\alpha}$ is the mole fraction of component $\alpha$ in the mixture, and $k_B$ is the Boltzmann constant.

To get an estimate of the configurational entropy as predicted by the pair correlation we rewrite $S_{c}$ in terms of the pair contribution to 
configurational entropy $S_{c2}$\cite{bssb},
\begin{equation}
S_{c}=S_{id}+S_{ex}-S_{vib}=S_{id}+S_{2}+\Delta S-S_{vib}=S_{c2}+\Delta S.
\label{sc2}
\end{equation}
\noindent
 Thus, the pair configurational entropy is defined as, 
 $S_{c2}=S_{id}+S_{2}-S_{vib}$. Note that an expansion of $S_{vib}$ in terms of two and many body would be interesting but to the best 
of our knowledge this has not been attempted before. Thus we have used total $S_{vib}$.

\subsection{ Mode coupling theory}
Many properties of  glass forming liquids can be explained by the well known mode coupling theory (MCT). This
microscopic theory can give a qualitative description of dynamical properties (such as temperature dependence of relaxation time) if the static
structure of the liquid is known and many experiments and simulation results have shown that MCT predictions hold good in the temperature regime of initial slow down
 of dynamics
\cite{Gotze,szamel-pre}.  
The equation for the intermediate scattering function for the binary mixture is given by,
\begin{widetext}
\begin{equation}
\ddot{\bm{S}}(k,t)+\bm{\Gamma} \dot{\bm{S}}(k,t)+{\bm{\Omega}}^2(k){\bm{S}}(k,t)+\bm{\Omega}^2(k)\int dt' \bm{\mathcal{M}}(t-t')\dot{\bm{S}}(k,t')=0,
\label{fkt-eqn}
\end{equation}
\end{widetext}
where $\bm{\Omega}^2(k)=\frac{k^2k_BT}{m}\bm{S^{-1}}(k)$ and ${\bm{S}}(k,t)$ is the matrix of intermediate scattering
functions $S_{\alpha \beta}(k,t)$ and the memory function $\bm{\mathcal{M}}$ can be written as :
\begin{widetext}  
\begin{eqnarray}
(\Omega^2(k)\mathcal{M}(k,t))_{\alpha \beta}=\frac{1}{2\rho \sqrt{x_\alpha x_\beta}}\sum_{ll'mm'}\int \frac{d\bm{q}}{(2\pi)^3}V_{\alpha lm}({\bf{q}},{\bf{k-q}})
                   V_{\beta l'm'}({\bf{q}},{\bf{k-q}}) \nonumber\\
                    \times S_{mm'}(\mid{\bf{k-q}}\mid)S_{ll'}(q)\phi_{mm'}(\mid{\bf{k-q}}\mid,t)\phi_{ll'}(q,t),
\label{memory-fkt}
\end{eqnarray}
\end{widetext}
where $\phi_{\alpha \beta}(k,t)=\frac{S_{\alpha \beta}(k,t)}{S_{\alpha \beta}(k,0)}$, ${\bf{k-q}}=\bf{p}$ and $ V_{\alpha lm}({\bf{q}},{\bf{p}})=[\hat{\bf{k}}.{\bf{q}}\delta_{\alpha m} C_{\alpha l}(q) + 
\hat{\bf{k}}.{\bf{p}}\delta_{\alpha l} C_{\alpha m}(p)]$ , here $\bm{C}(q)$ is defined as $\bm{S^{-1}}(q)=\bm{1}-\bm{C}(q)$.
 The static structure factor is obtained from 
computer simulation.

   Solving Eq.\ref{fkt-eqn} we have calculated the relaxation times, $\tau_{\alpha \beta}$ from
$\phi_{\alpha \beta}(q,t)$ at $1/e$ and the average relaxation time is obtained from the following equation,
\begin{equation}
 \tau_{MCT}=\sum_{\alpha \beta}x_{\alpha}x_{\beta}\tau_{\alpha \beta}.
\label{tau-mct}
\end{equation}

\section{Result and discussion}
In this section using simulation results,
 we discuss the role of entropy in determining the dynamics via the Adam Gibbs relationship.
We also investigate the role of pair correlation function in the dynamics by calculating the pair configurational entropy from
radial distribution function. All these studies are done for LJ and WCA systems at different densities.

\subsection{Adam Gibbs relation}
\begin{figure}[h]
\centering
\subfigure{
\includegraphics[width=0.45\textwidth]{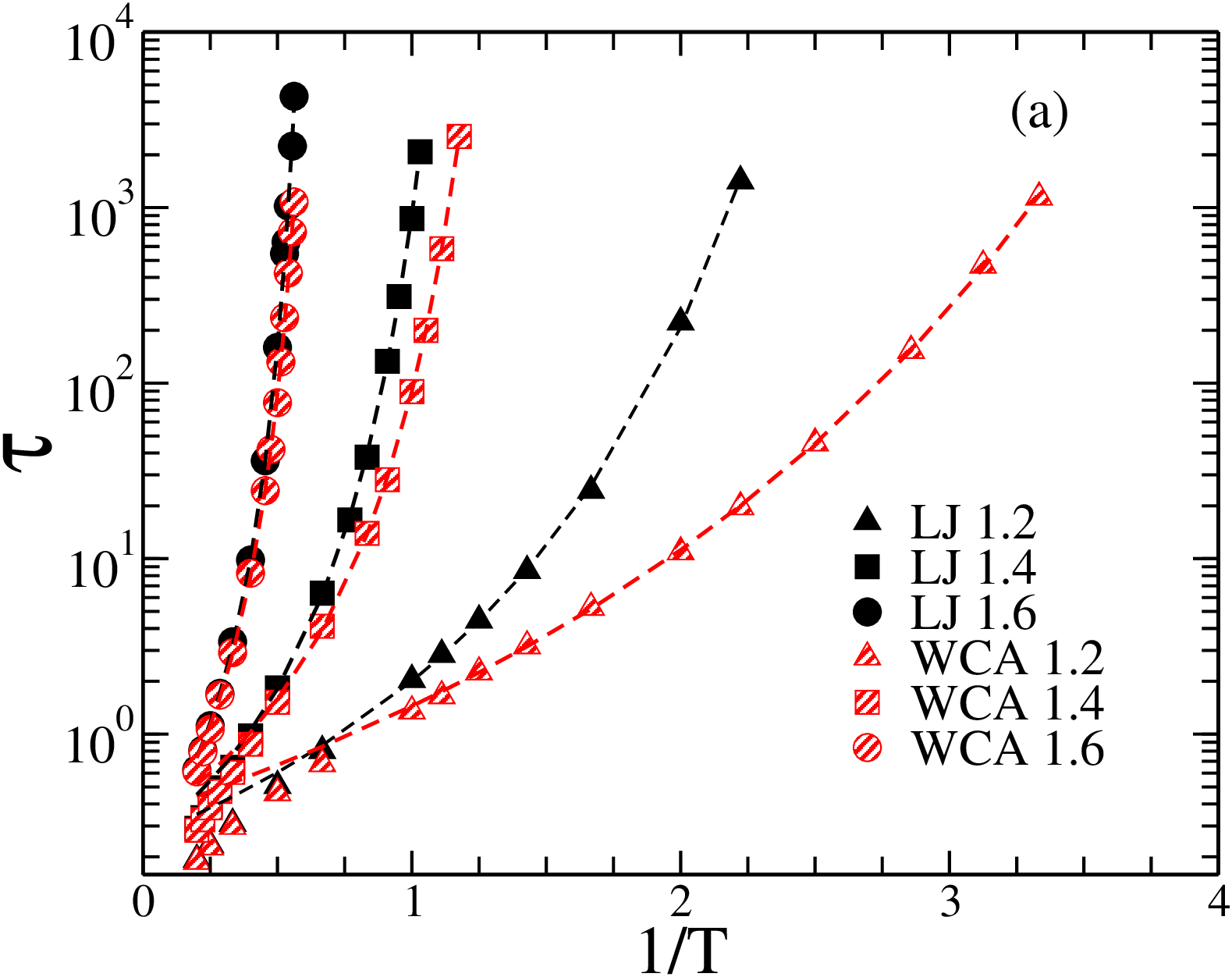}}
\subfigure{
\includegraphics[width=0.45\textwidth]{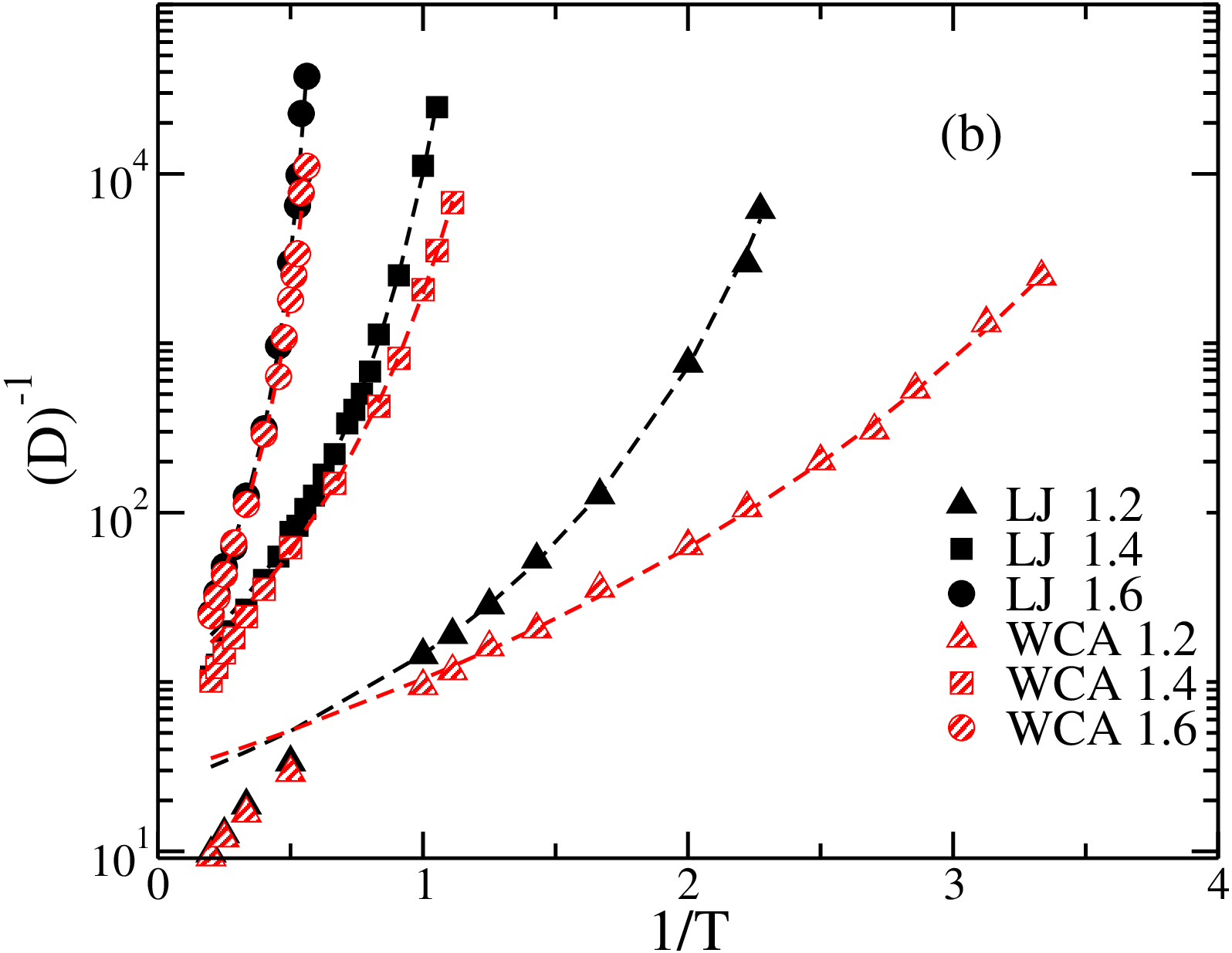}}
\caption{(a)The VFT fit  of the relaxation time $\tau$ defined from overlap function $<q(t=\tau)>= 1/e$. The plot shows LJ system is more fragile than WCA system. 
At higher densities the difference in dynamics between LJ and WCA systems decreases. (b) The VFT fit of inverse of diffusion coefficient shows qualitatively similar
behaviour as found for relaxation time. The values of kinetic fragilities from both the quantities have been given in the Table \ref{kinetic-fragility-table}. }
\label{kinetic-fragility-plot}
\end{figure}

The relaxation time and diffusion coefficient obtained from simulation can be well fitted to VFT form using divergence temparature,
$T_{VFT}$ and kinetic fragility, $K_{VFT} $  as fitting parameters. In Fig.\ref{kinetic-fragility-plot} we show that within the density range studied here,
  the LJ system is more fragile than the WCA system.  
The difference between kinetic fragility for LJ and WCA system is maximum at $\rho =1.2$ and as density increases
the difference decreases. In Table \ref{kinetic-fragility-table} the data shows that similar to $K_{VFT}$, the divergence temperature $T_{VFT}$ increases with density 
and the difference between $T_{VFT}$ values for LJ and WCA systems reduces as density increases.

\begin{table}[h]
\caption{ The values of kinetic fragility ($K_{VFT}$) and  the divergence temperature ($T_{VFT}$) for all three densities are tabulated below.
 At higher density both the values of $K_{VFT}$ and $T_{VFT}$ come closer for LJ and WCA  systems which indicate that the dynamical differences
between the systems decrease at higher density .} 
 \begin{tabular}{|c|c|c|c|c|}
 \hline
    $\rho$ & $K_{VFT}^{LJ}$  &$K_{VFT}^{WCA}$ & $K_{VFT}^{LJ}$  &$K_{VFT}^{WCA}$  \\ 
 & from $\tau$ & from $\tau$&from D & from D\\
\hline
    1.2 &0.187& 0.124 &0.151&0.116\\ \hline
    1.4 & 0.255& 0.237&0.209&0.166\\ \hline
    1.6 &0.291&0.285&0.284&0.257\\ \hline
\end{tabular}
 \begin{tabular}{|c|c|c|c|c|}
    \hline
    $\rho$ & $T_{VFT}^{LJ}$  &$T_{VFT}^{WCA}$ & $T_{VFT}^{LJ}$  &$T_{VFT}^{WCA}$  \\ 
  &from $\tau$ & from $\tau$&from D & from D\\
\hline
  1.2 & 0.278& 0.150 &0.239&0.132 \\ \hline
  1.4 &  0.678&0.579 &0.592&0.471\\ \hline
  1.6 &  1.319&1.258&1.255&1.153\\ \hline
\end{tabular}
\label{kinetic-fragility-table}
\end{table}

\begin{figure}[h]
\centering
\subfigure{
\includegraphics[width=0.45\textwidth]{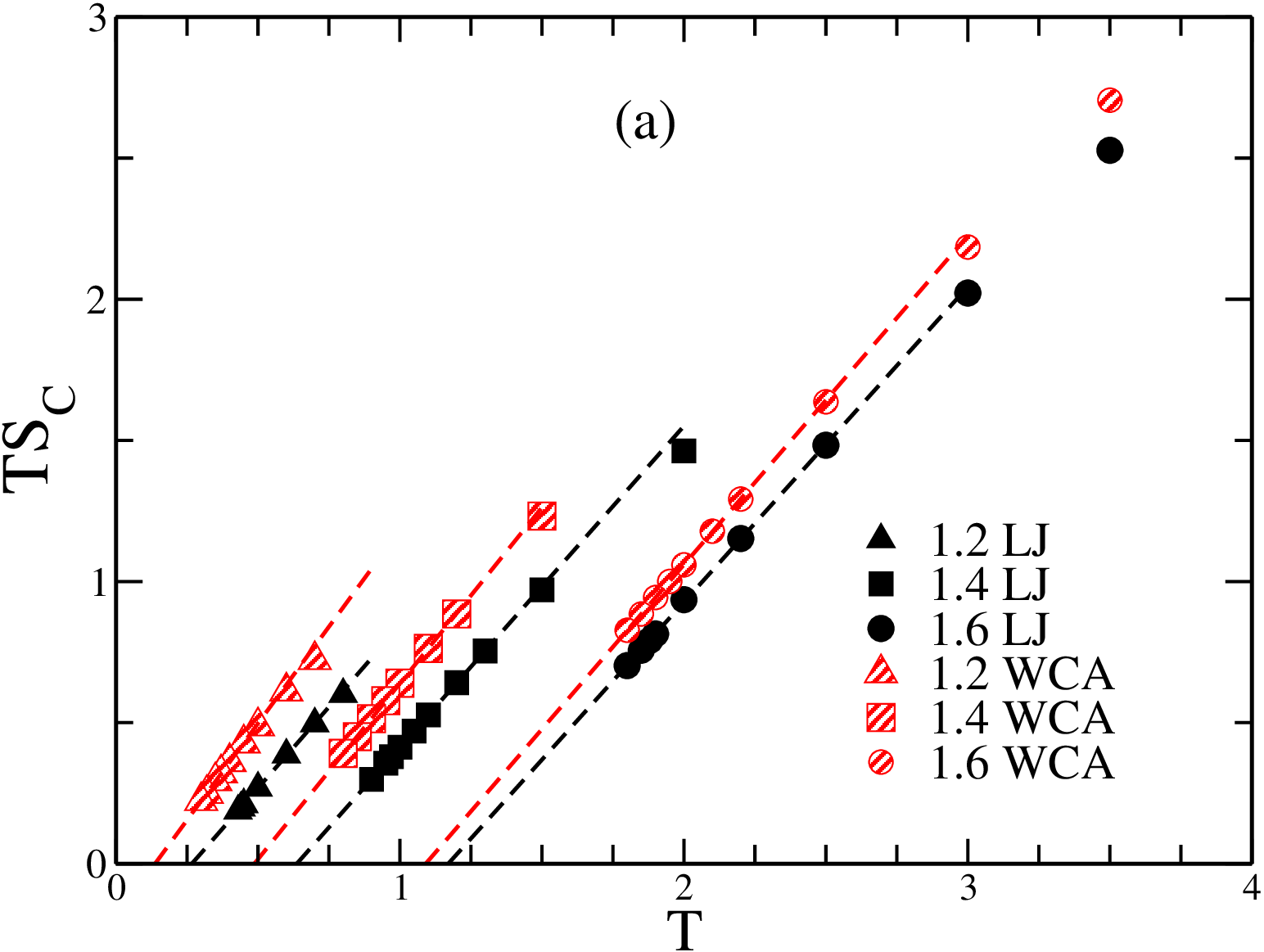}}
\subfigure{
\includegraphics[width=0.45\textwidth]{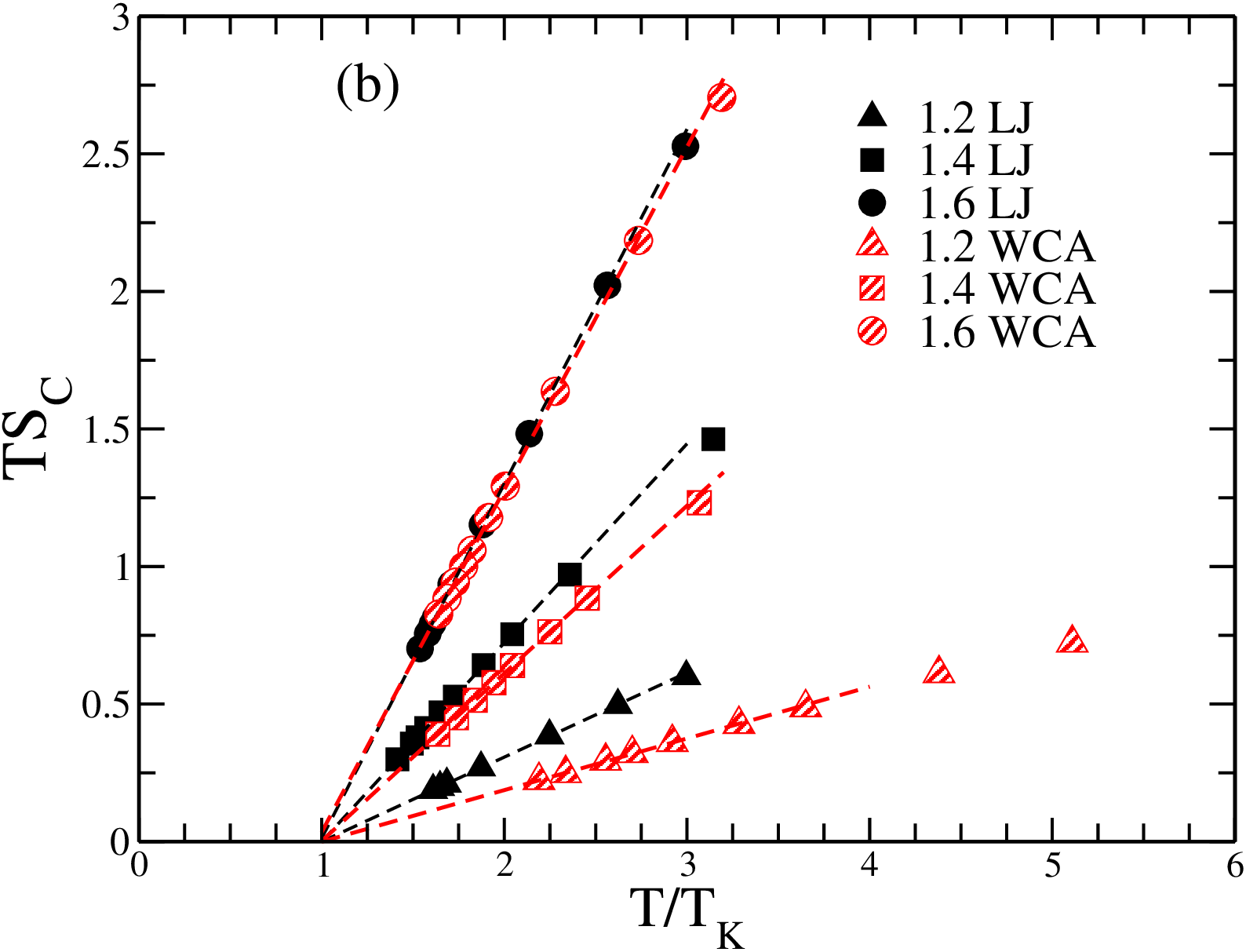}}
\caption{ (a) Temperature dependence of $TS_c$ for the studied models to determine
the Kauzmann temperature.
(b) $TS_c$ vs. $T/T_K$ for all the densities for both the systems. The thermodynamic fragility, $K_T$, is determined from the slope of the plot.
 $T_K$ and $K_T$ values are given in Table \ref{thermodynamic-fragility-table} for all the systems. Both the plot show that
  at high densities the difference between the LJ and WCA systems decrease. }
\label{thermodynamic-fragility-plot}
\end{figure}

In an earlier study we have shown that the kinetic fragility both for LJ and WCA systems at $\rho=1.2$ is related to the thermodynamic fragility \cite {bssb}. 
In a similar fashion, here  we analyze the relationship between kinetic and thermodynamic fragility over a wider range of densities.
 We compute  configurational entropy at different densities. Thermodynamic integration is used to calculate the total entropy and subtracting the vibrational
entropy from it (details given in Sec. IIID) we obtain configurational entropy. For all the systems studied, $TS_c$ show a linear behaviour when plotted with temperature.
 We calculate the Kauzmann temperature, $T_K$,
by extrapolation of this plot (Fig.\ref{thermodynamic-fragility-plot}). The values are similar to the corresponding $T_{VFT}$ values (Table \ref{thermodynamic-fragility-table}).
 Further, the thermodynamic fragility $K_T$, 
which is the slope of $TS_c$ vs. $T/T_K$ plot,
can be computed. We find that $K_T$ has a strong density dependence and increases with density.
However similar to that found in the case of kinetic fragility, the difference between LJ and WCA systems in
terms of thermodynamic fragility also decreases at higher densities (Table \ref{thermodynamic-fragility-table}).

\begin{table}[h]
\caption{ The values of thermodynamic fragility and  Kauzmann temperature  for all three densities are tabulated below. At higher density both the values of $K_T$ and $T_K$  
come closer for the LJ and the WCA 
systems. AG coefficients (both from relaxation time, $A^\tau$ and diffusion coefficient, $A^D$) for LJ and WCA systems have also been tabulated here. The AG coefficient increases with increase of density.
The values of $K_{AG}=K_{T}/A^\tau$ are also tabulated here.} 
 \begin{tabular}{|c|c|c|}
    \hline
    $\rho$ & $K_T^{LJ}$  &$K_T^{WCA}$  \\ \hline
    1.2 &0.3084&0.1792 \\ \hline
    1.4& 0.723& 0.609\\ \hline 
    1.6&1.307&1.270\\   \hline 
\end{tabular}
 \begin{tabular}{|c|c|c|}
    \hline
     $T_K^ {LJ}$  &$T_K^{WCA}$  \\ \hline
    0.27 &0.134  \\ \hline
    0.631 & 0.50  \\ \hline
     1.18&1.11\\ \hline
\end{tabular}
\begin{tabular}{|c|c|c|c|c|}
    \hline
   $\rho$ & $A_{LJ}^\tau$ & $A_{WCA}^\tau$& $A_{LJ}^D$& $A_{WCA}^D$\\ \hline

  1.2&  1.87 & 1.89 & 1.50&1.51 \\ \hline
  1.4 &  3.57 & 4.37&2.63 &3.30 \\ \hline
  1.6 &   6.96&7.57& 5.45 &5.84 \\ \hline
\end{tabular}
 \begin{tabular}{|c|c|c|}
    \hline
    $K_{AG}^{LJ}$  &$K_{AG}^{WCA}$  \\ \hline
    0.165&0.095\\ \hline
    0.202&0.139 \\ \hline 
    0.187&0.167\\ \hline
\end{tabular}
\label{thermodynamic-fragility-table}
\end{table}

\begin{figure}[h]
\centering
\subfigure{
\includegraphics[width=0.45\textwidth]{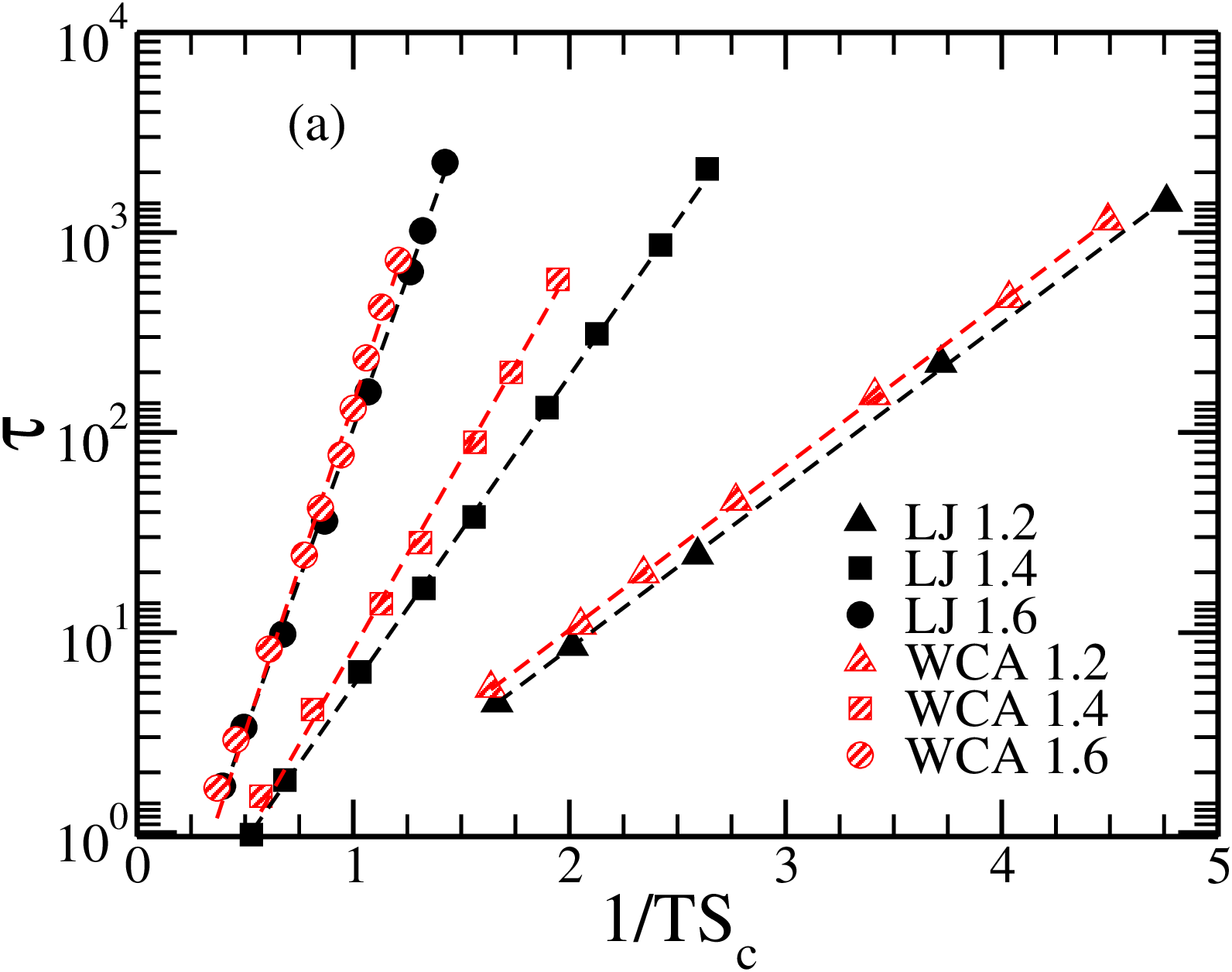}}
\subfigure{
\includegraphics[width=0.45\textwidth]{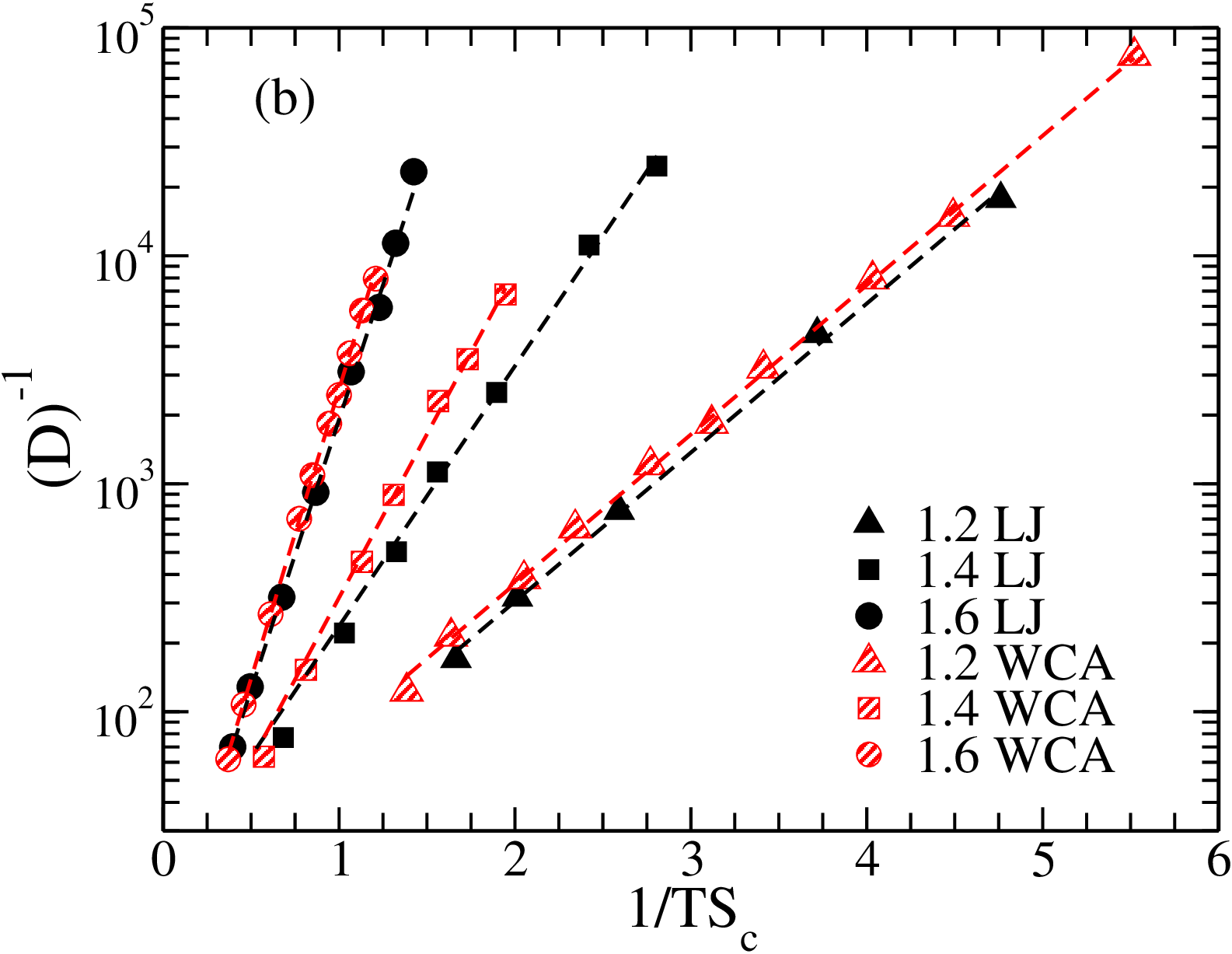}}
\caption{ The Adam-Gibbs plot in terms of (a) relaxation time and (b) diffusion coefficients. Both the plots show that the
temperature dependence of dynamical quantities are quantitatively captured by the temperature variation of the configurational entropy.  }
\label{adam-gibbs-plot}
\end{figure}
 Next we study the validity of  the Adam Gibbs relation which connects the thermodynamics to the dynamics and find it to be valid for all the systems studied here. 
 Fig.\ref{adam-gibbs-plot} shows the Adam-Gibbs plot  both for $\tau$ and $D^{-1}$.  Thus our study shows that  the temperature variation of the configurational entropy $S_c$  
captures the dynamics and also the differences in the dynamics between the LJ and the WCA systems. The Adam Gibbs coefficient (A) also has a strong density dependence
and increases with density (Table \ref{thermodynamic-fragility-table}). An earlier study has shown that the Adam Gibbs coefficient (A) is  related to the
 high temperature activation energy (E) \cite{shila-jcp}.
It has also been found that
 `E' increases with increase in density \cite{tarjus_pre,unravel}. Thus our finding of increase in `A' value with 
density supports the earlier studies.

\begin{figure}[h]
\centering
\includegraphics[width=0.45\textwidth]{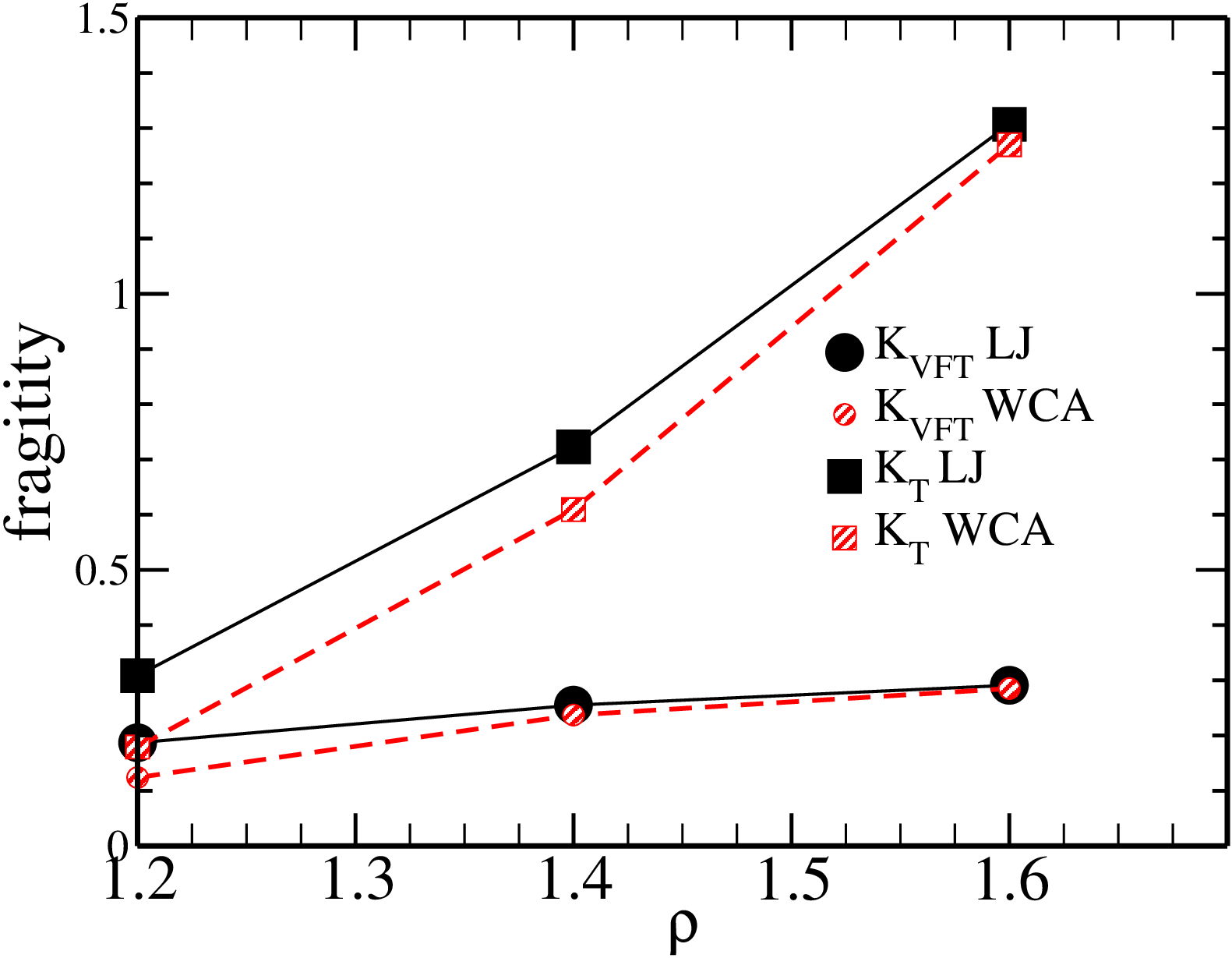}
\caption{Density dependence of thermodynamic and kinetic fragilities. Thermodynamic fragilities show strong temperature dependence, kinetic fragilities show weak temperature dependence for both LJ and WCA systems.}
\label{kinetic-thermodynamic-fragilities}
\end{figure}

 Note that we also find that although the thermodynamic fragility and the Adam Gibbs coefficients have strong density dependence, their ratio (Table \ref{thermodynamic-fragility-table}) 
which is related 
to kinetic fragility (see Sec. IIIC) shows a weak  dependence on density. This kind of observation on density dependence has already been reported for the LJ system
\cite{tarjus_berthier_jcp,shila-epje}. However 
although unlike the  LJ system, the WCA system does not show density-temperature scaling and thus density independence of fragility \cite{tarjus_berthier_jcp}, compared to the 
thermodynamic fragility the density dependence of the kinetic fragility even in WCA system is found to be weak (Fig.\ref{kinetic-thermodynamic-fragilities}).

Note that from Fig.\ref{thermodynamic-fragility-plot}a we find that at same temperature, the configurational entropy is different for the two systems.
Since $S_c$ has two different components, the total entropy and the vibrational entropy, it would be interesting to understand which part of the entropy gives rise to
the difference in $S_c$.

\subsection{Density of states and vibrational entropy }
\begin{figure}[h]
\centering
\subfigure{
\includegraphics[width=0.45\textwidth]{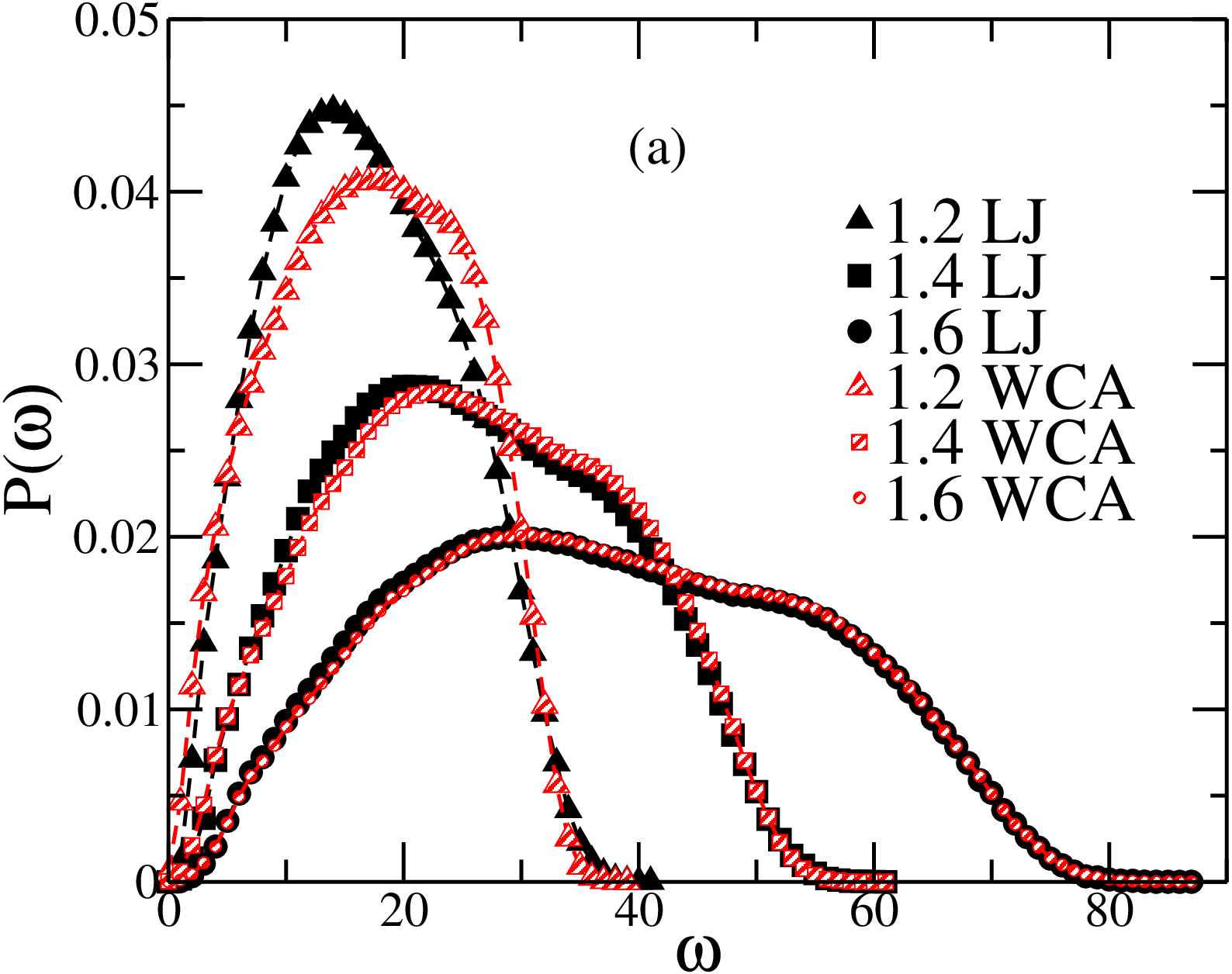}}
\subfigure{
\includegraphics[width=0.45\textwidth]{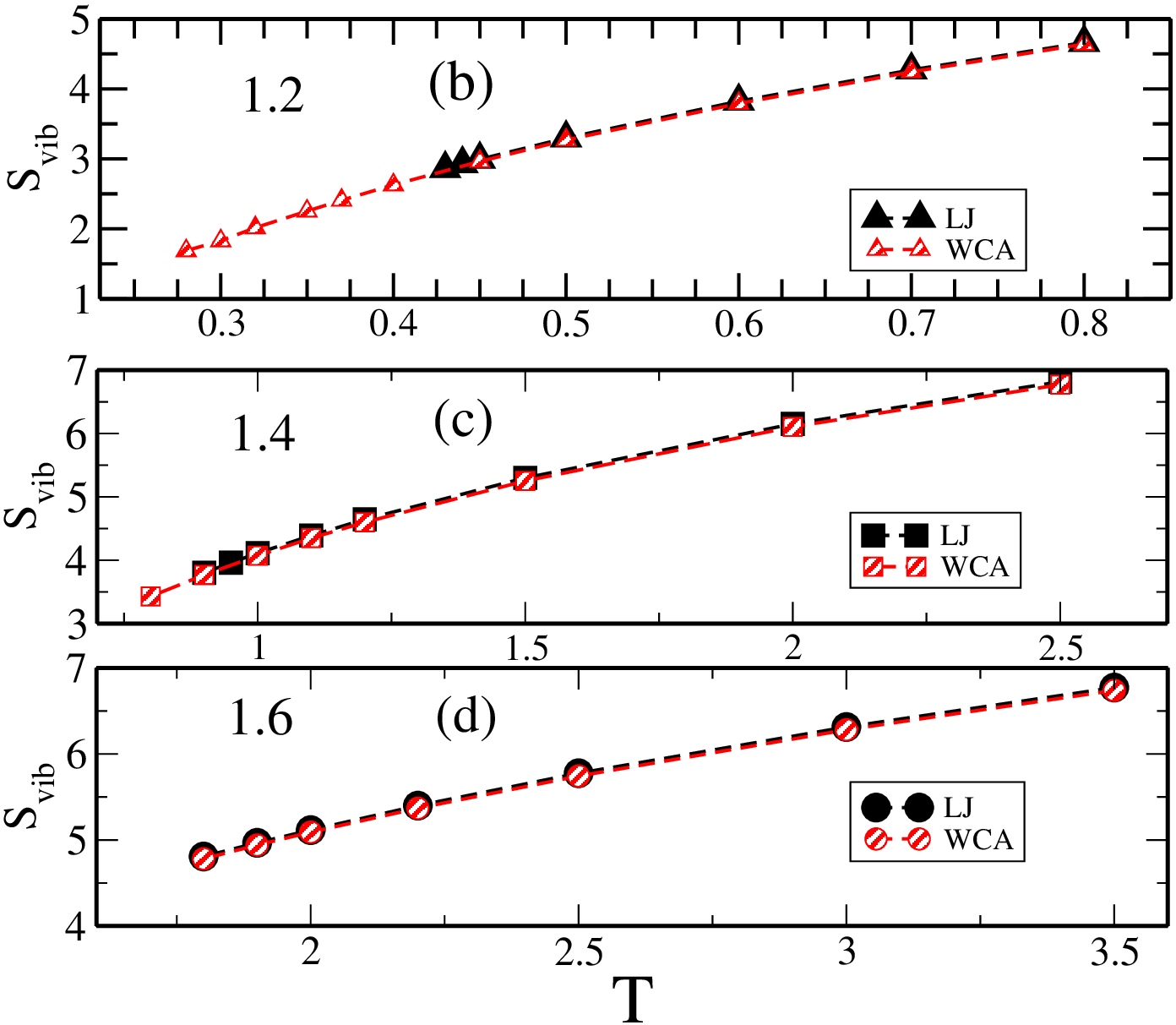}}
\caption{ (a) Distribution of density of vibrational states , $P({\omega}$), for 1.2,1.4,1.6 density (T=0.45,1.00,1.90 respectively).
 At $\rho=1.2$ the difference in the distribution of 
$P({\omega}$) for LJ and WCA system is maximum. At higher density the distributions are almost overlapping for LJ and WCA systems.
The temperature dependence of vibrational entropy for (b) 1.2, (c)1.4, (d)1.6 density. $S_{vib}$ hardly shows any difference between LJ and WCA systems.}
 

\label{DOS_fig}
\end{figure}

In this section, we analyse the vibrational density of states of the LJ and the WCA systems.
The vibrational entropy is calculated from vibrational density of states (VDOS), $P(\omega)=(1/3N)<\sum_l\delta(\omega-\omega_l)>$
, where $\omega_l$ is the frequency of the mode l. Normal modes of vibration are obtained by diagonalizing the
Hessian matrix using LAPACK \cite{lapack}. We find that similar to what has been reported earlier \cite{Wang-Xu-prl}, the difference between VDOS for the LJ and WCA 
systems are maximum at $\rho=1.2$ and as density
increases they come closer,
as  a result of which at $\rho =1.6$, VDOS more or less overlap (Fig.\ref{DOS_fig}a). However in the calculation of $S_{vib}$ we find that the difference between the two systems is negligible
through out the density range studied here. At $\rho=1.2$ although VDOS are different, because of the cancellation of the values at low and high frequency of VDOS the $S_{vib}$
appears to be similar. The WCA system has more low frequency modes and at 
$\rho =1.2$ we find some rattlers with zero frequency modes whose contribution has been ignored in the present study.

\begin{figure}[h]
\centering
\subfigure{
\includegraphics[width=0.35\textwidth]{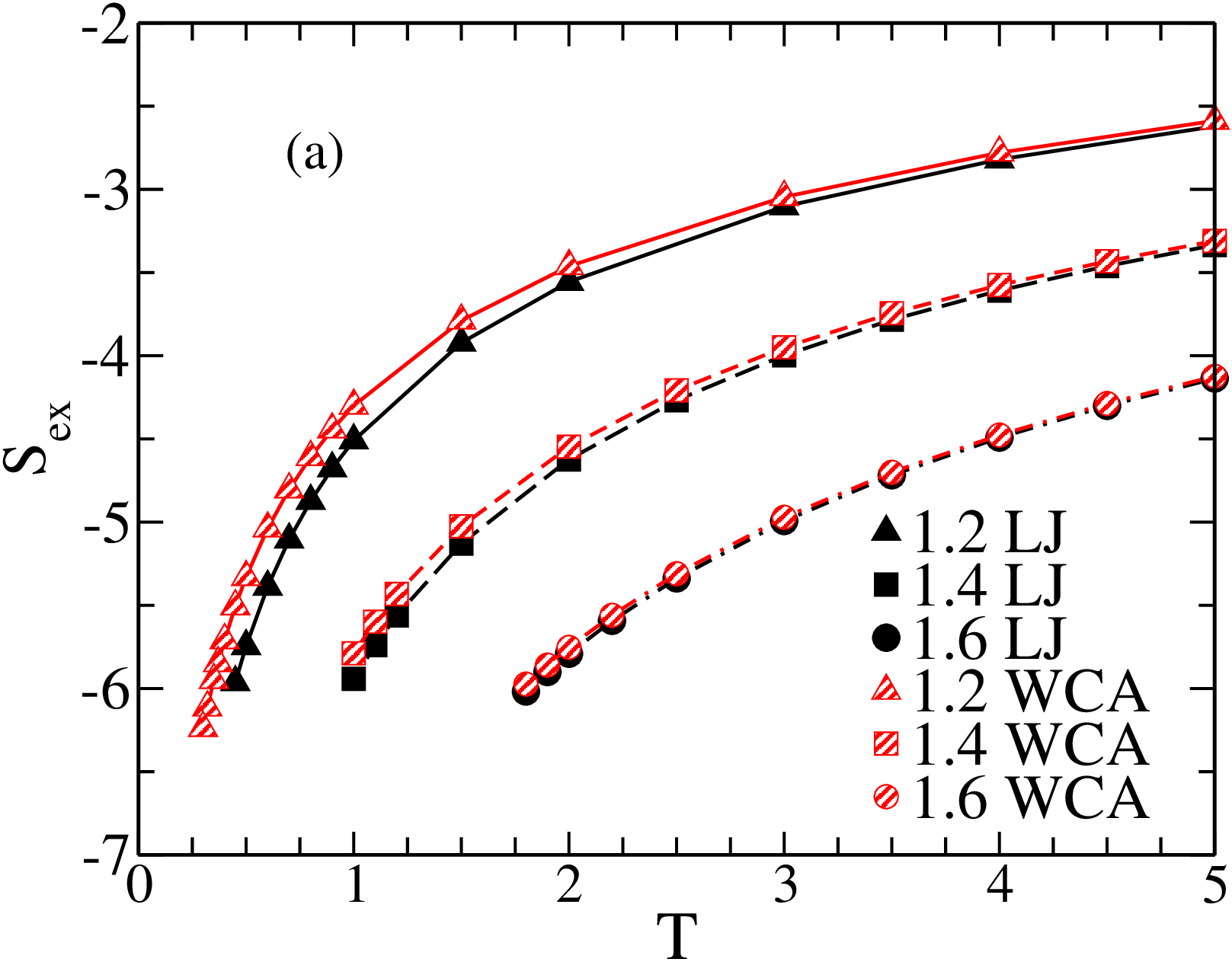}}
\subfigure{
\includegraphics[width=0.35\textwidth]{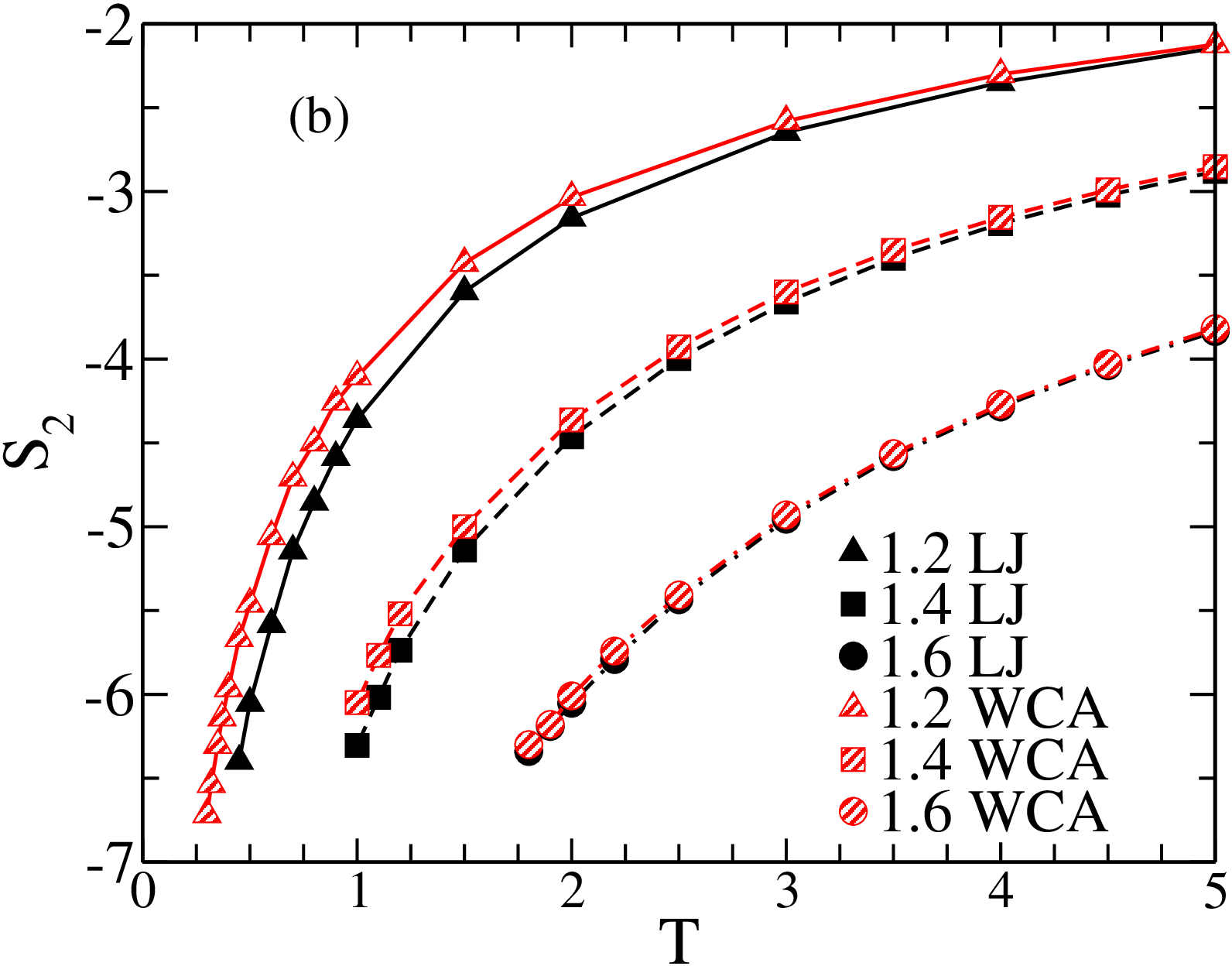}}
\subfigure{
\includegraphics[width=0.35\textwidth]{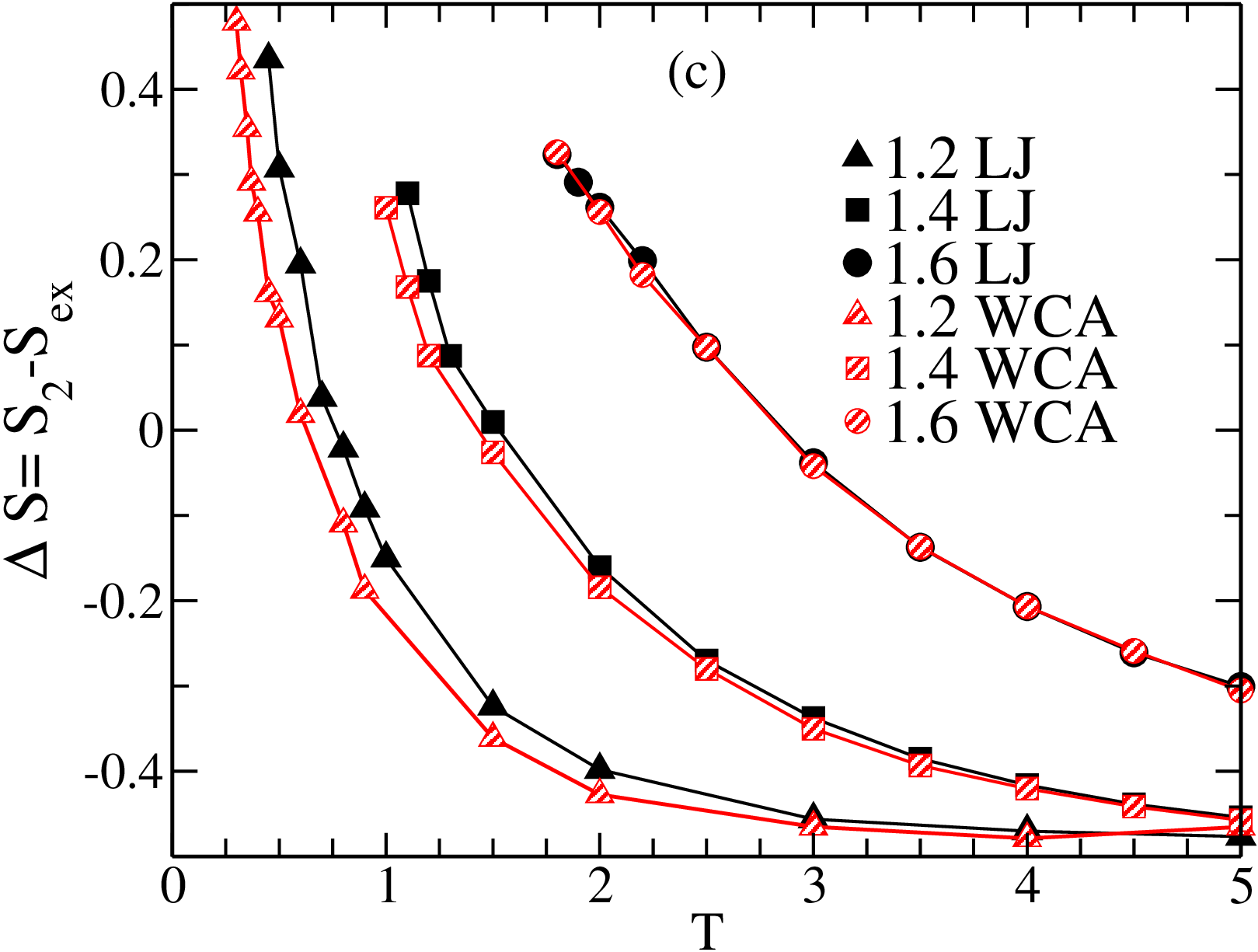}}
\caption{(a) $S_{ex}$, calculated from thermodynamic integration, with temperature for LJ and WCA systems at different densities. (b) Pair entropy ($S_2$) calculated from
pair correction function using Eq.\ref{s2_final} for all the systems. Both $S_{ex}$ and $S_{2}$ have negative values and decreases at lower temperature. (c) The temperature
variation of  residual multiparticle entropy ($\Delta S$). All the plots show that at  higher densities the difference between LJ and WCA systems decrease. }
\label{S2_Sex_deltaSfig}
\end{figure}

\subsection{ Entropy : pair and higher order }

In the above section, we have shown that $S_{vib}$ is similar for LJ and WCA systems in all three densities.
This implies that the difference in $S_c$ comes from the total entropy. 
In that, the $S_{id}$  is same for the LJ and WCA systems at a fixed density and temperature, thus 
it is the excess entropy which is responsible for the difference. Fig.\ref{S2_Sex_deltaSfig}a shows indeed the $S_{ex}$ of the two systems are different and as density increases the 
difference in $S_{ex}$ between LJ and WCA systems decreases. This observation again supports our finding that at higher density both the systems thermodynamically behaves 
in a similar way.

    As given in Sec. IIIE, the $S_{ex}$ can be separated into pair and higher order terms. We find that both the $S_{2}$ and $\Delta S$ are different for
 LJ and WCA systems.
As it has been already shown that the higher order terms are different for the two systems \cite{coslovich}, thus
the difference in $\Delta S$ is not unexpected. However since the pair structures are similar, the difference in pair entropy is not intuitive.
But our study shows that the difference in $S_{ex}$ is driven by $S_2$ as the 
contribution from $\Delta S$ is really small (note  the y-axis scale of Fig.\ref{S2_Sex_deltaSfig}b and Fig.\ref{S2_Sex_deltaSfig}c).
 We find that similar to 
$S_{ex}$, the maximum difference in $S_{2}$ is present at 1.2 density.
As density increases the difference between the two systems in $S_2$ and $\Delta S$ decreases.

\begin{figure}
\centering
\subfigure{
\includegraphics[width=0.35\textwidth]{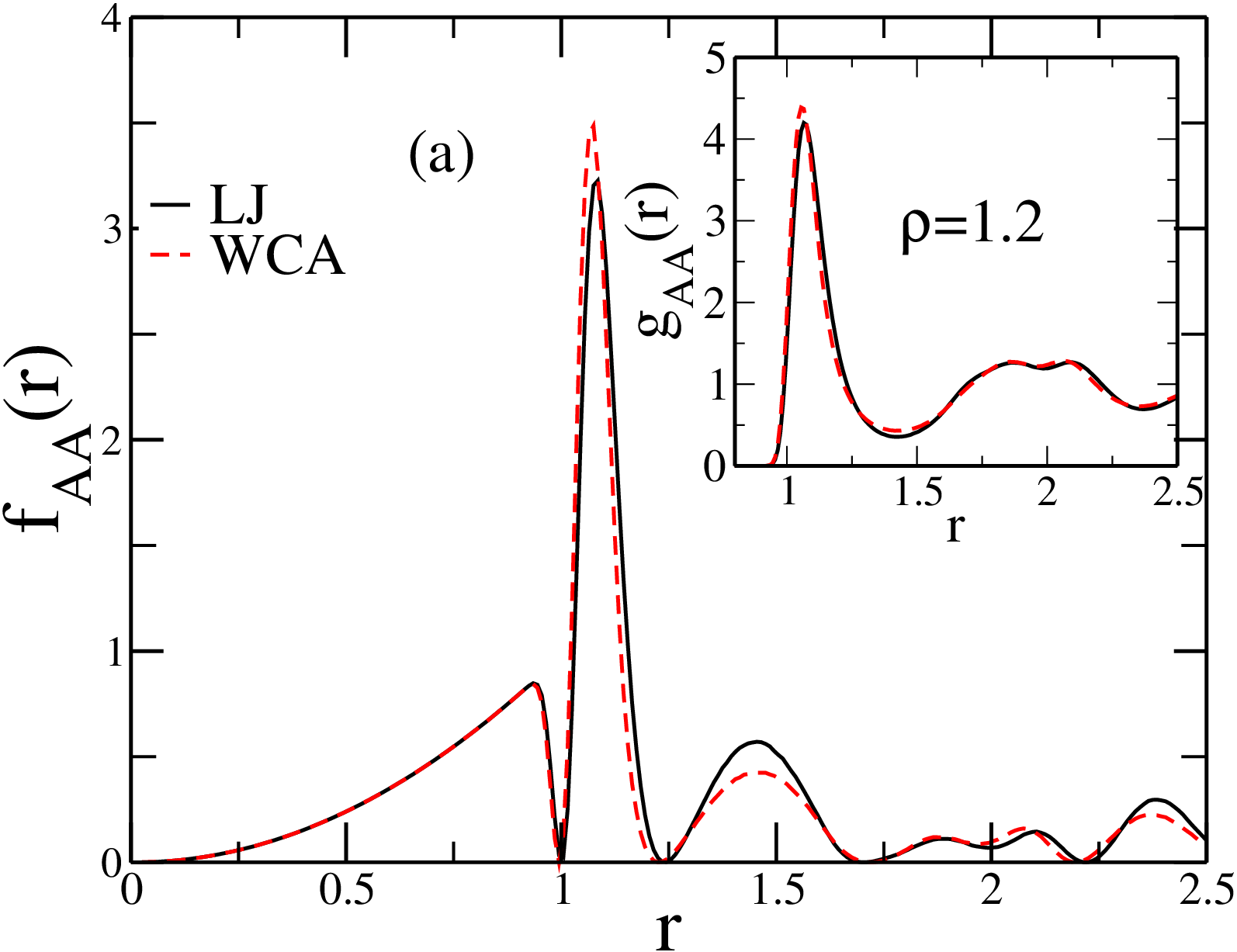}}
\subfigure{
\includegraphics[width=0.35\textwidth]{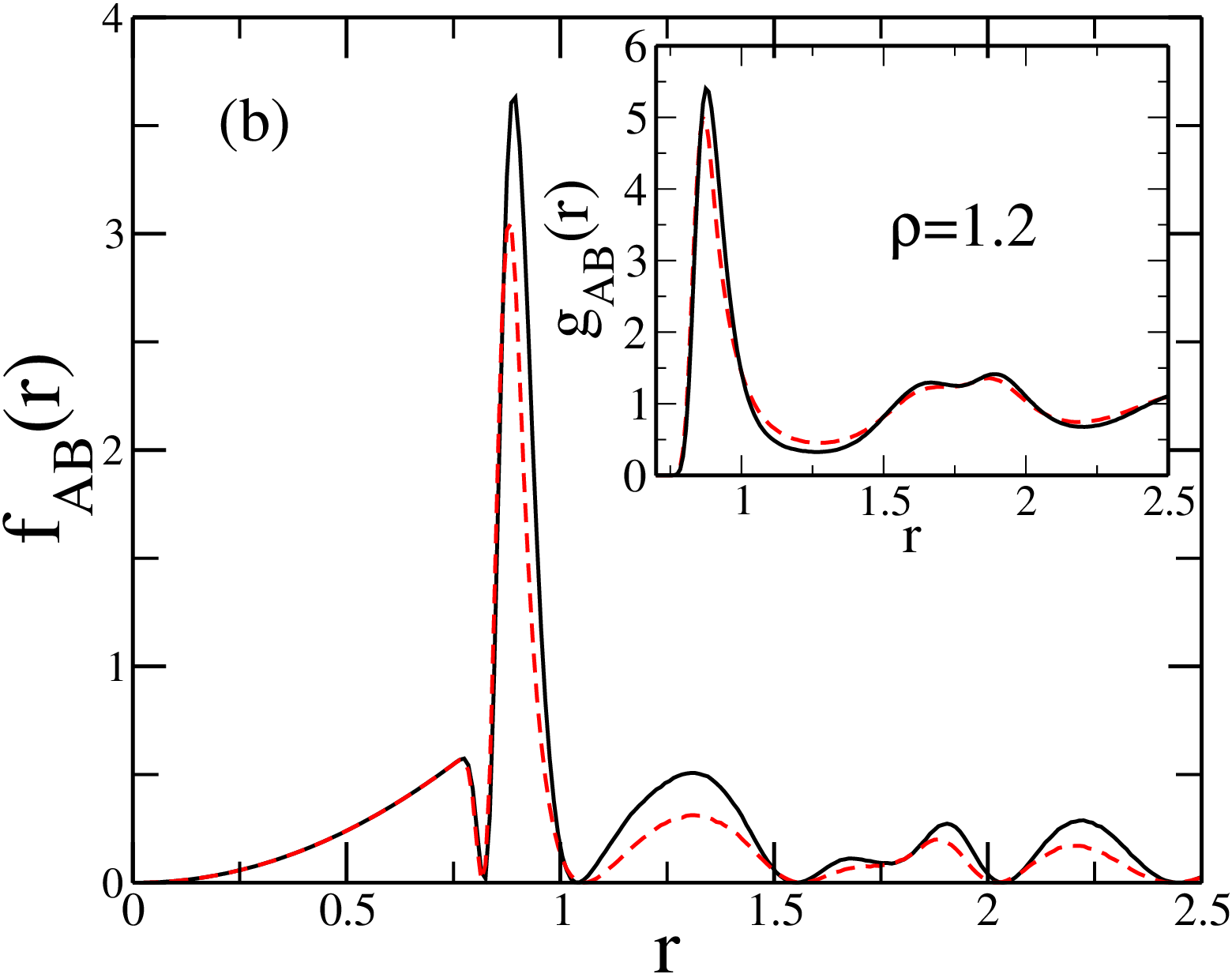}}
\subfigure{
\includegraphics[width=0.35\textwidth]{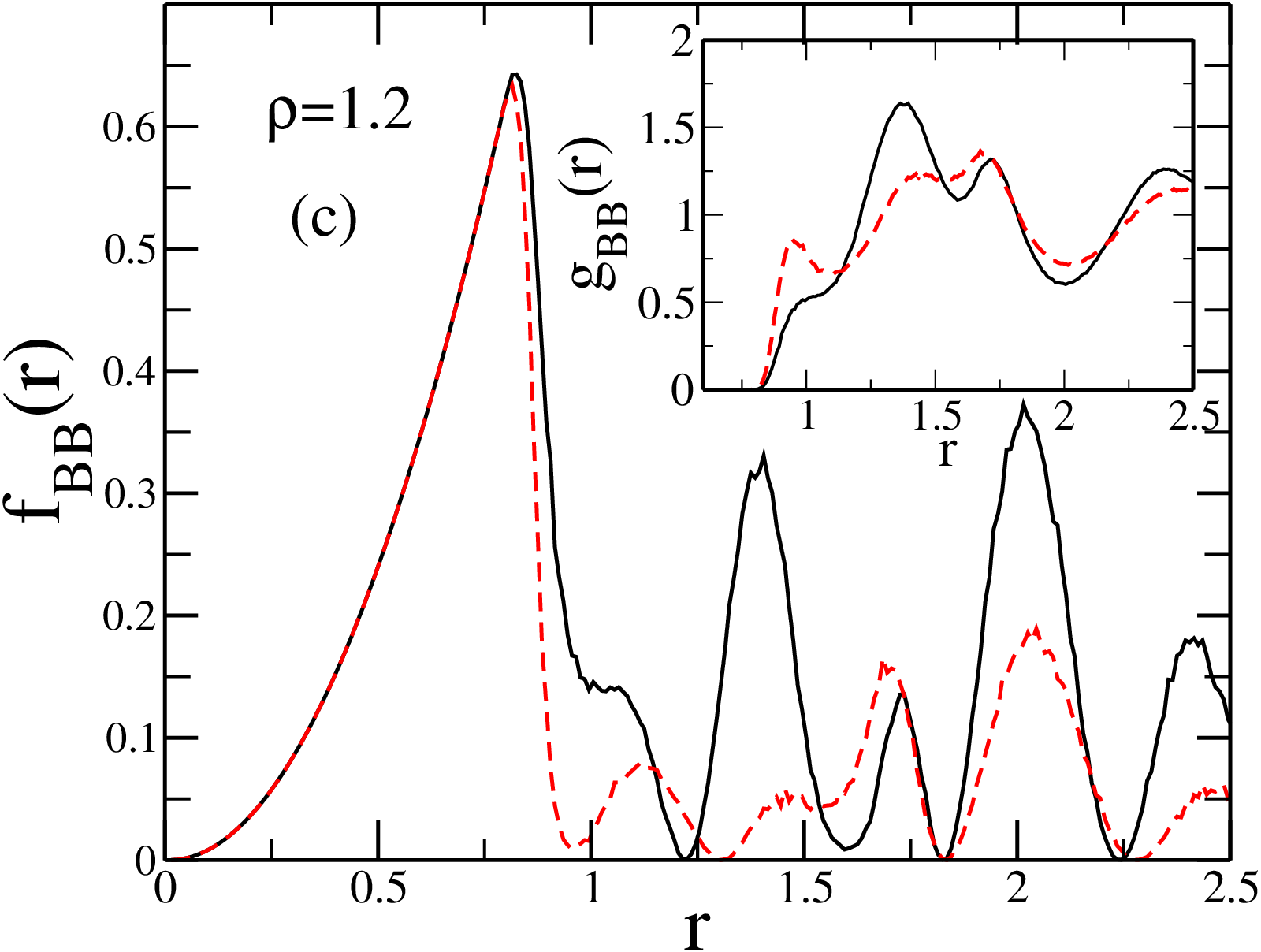}}
\subfigure{
\includegraphics[width=0.35\textwidth]{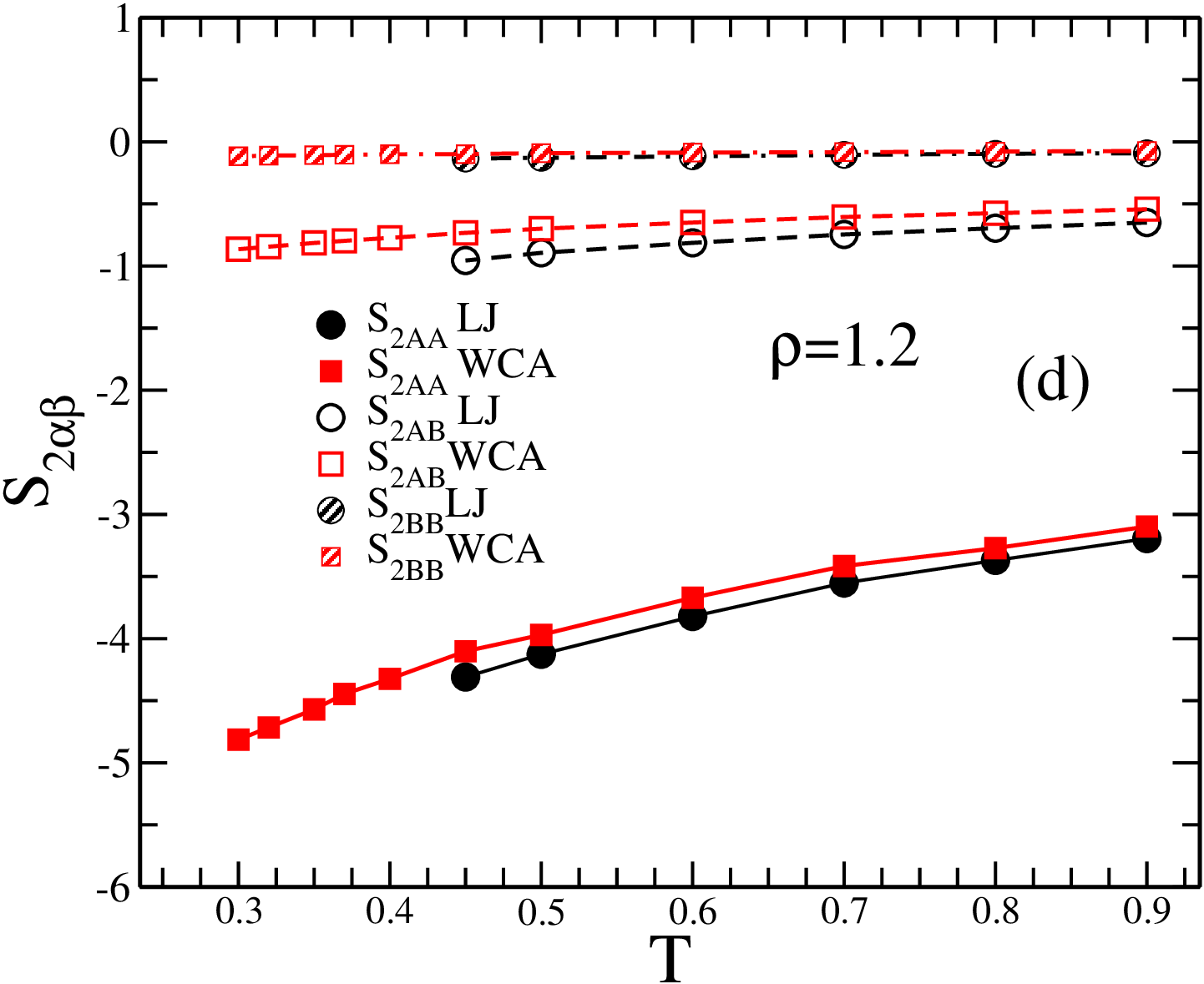}}
\caption{
 (a) Comparison of the `r' dependence of $f_{AA}(r)$ (Eq.\ref{fr_eq}) for LJ and WCA systems.
(inset) Radial distribution function for A-type particles. (b) Similar plot for AB and (c) BB pair. All the plots show that 
the small difference in rdf peak and minima position
amplify in the entropic terms ($f_{\alpha\beta}(r)$). (d) The partial contribution of AA, AB, BB type particles to the total entropy.
 This plot shows that although rdf and $f_{BB}(r)$ for B-type particles
have maximum difference between LJ and WCA systems, the difference in pair  entropy comes from AA and AB pairs.
All the plots are given for $\rho=1.2$.  }
\label{S2_part_by_part}
\end{figure}

 Next we analyse the origin of the difference in $S_2$ of the two systems. Although the total radial distribution function
 of LJ and WCA systems look similar \cite{tarjus_prl},
in Fig.\ref{S2_Sex_deltaSfig}b we find that pair entropy
for LJ and WCA systems are quite different and  the difference is maximum at $\rho=1.2$. 
However it has been earlier shown that  
 the partial radial distribution functions are not identical at this density \cite{Pedersen-prl} and the radial distribution function
 of the B-type particles are indeed quite different (Fig.\ref{S2_part_by_part}c inset). So the difference in pair entropy is expected to arise from the rdf of B-type particles.
Also it is imperative to understand if the difference comes from the short or long range part of the rdf.
As the difference in $S_2$ is maximum at $\rho=1.2$, we choose this density for our next analysis and address these two points mentioned above. 

The pair entropy $S_{2}$ for a binary system can be written in terms of 
the partial radial distribution functions,
\begin{equation}
\frac{S_{2}}{k_{B}}=\sum_{\alpha,\beta} S_{2\alpha\beta}=-2\pi\rho \sum_{\alpha,\beta}x_{\alpha} x_{\beta} \int_0^{\infty} f_{\alpha\beta}(r) d {r},
\label{s2}
\end{equation}
\noindent where, 
\begin{equation}
 f_{\alpha\beta}(r)=\{g_{{\alpha}{ \beta}}(r) \ln g_{{\alpha} {\beta}}(r)- [g_{{\alpha}{\beta}}(r)-1]\} r^2.
\label{fr_eq}
\end{equation}
\noindent  Here $ g_{{\alpha}{ \beta}}(r)$ is the atom-atom pair correlation between atoms of type $\alpha$ and $\beta$, $\rho$ is the density of the system,
 $x_{\alpha}$ is the mole fraction of component $\alpha$ in the mixture, and $k_B$ is the Boltzmann constant.

$S_{2\alpha\beta}$ is the partial  pair entropy for a binary system given by,
\begin{equation}
\frac{S_{2\alpha\beta}}{k_{B}}=-2\pi\rho x_{\alpha} x_{\beta} \int_0^{\infty} f_{\alpha\beta}(r) d {r}.
\label{s2}
\end{equation}

 The plot of $f_{\alpha\beta}(r)$ vs. r  shows that the difference mainly comes from the first few peak and minima positions of the  rdf and not from large r
(Fig.\ref{S2_part_by_part}a, Fig.\ref{S2_part_by_part}b, Fig.\ref{S2_part_by_part}c).
We also find that although $g_{AA}(r)$ and $g_{AB}(r)$ (inset of Fig.\ref{S2_part_by_part}a, Fig.\ref{S2_part_by_part}b) for both the systems are very similar,
 the entropic terms ($f_{\alpha\beta}$(r)) for AA and AB pairs show enough
difference.
 More interestingly,
 we find that although
as a reflection of $g_{BB}(r)$,
 the $f_{\alpha\beta}(r)$ of B-B type of
particles looks quite apart,  but it has negligible effect on $S_{2BB}$. The small mole fraction of B-type of particles,
minimizes the differences (Fig.\ref{S2_part_by_part}d). The primary contribution to the difference in $S_2$ between the LJ and WCA systems come from
$S_{2AA}$ and $S_{2AB}$. Thus, our study shows that small difference at the two body level is  properly amplified in $S_2$, which allows us to disentangle the LJ and WCA systems.

Next to calculate the effect of pair correlation on the entropy, we consider the separation of
 the configurational entropy into pair and many body parts as described earlier (Sec. IIIE)\cite{bssb}.
Using $S_2$ information we calculate the pair configurational entropy $S_{c2}$, which goes to zero at the Kauzmann temperature ($T_{K2}$)
 given in the Table \ref{pair-thermodynamic-fragilities-table}.
 It has been earlier
reported that this temperature has a connection to the MCT transition temperature \cite{unravel}. The pair thermodynamic fragility  $K_{T2}$
 is computed from the slope of $T_{Sc2}$ vs. $T/T_{K2}$ plot. The plots show that the thermodynamic
fragility even at the two body level is higher for the LJ system. Like relaxation time and thermodynamic fragility  we find that the difference in pair thermodynamic
 fragility between LJ and WCA systems also decreases as density increases.

\begin{figure}[h]

\centering
\includegraphics[width=0.5\textwidth]{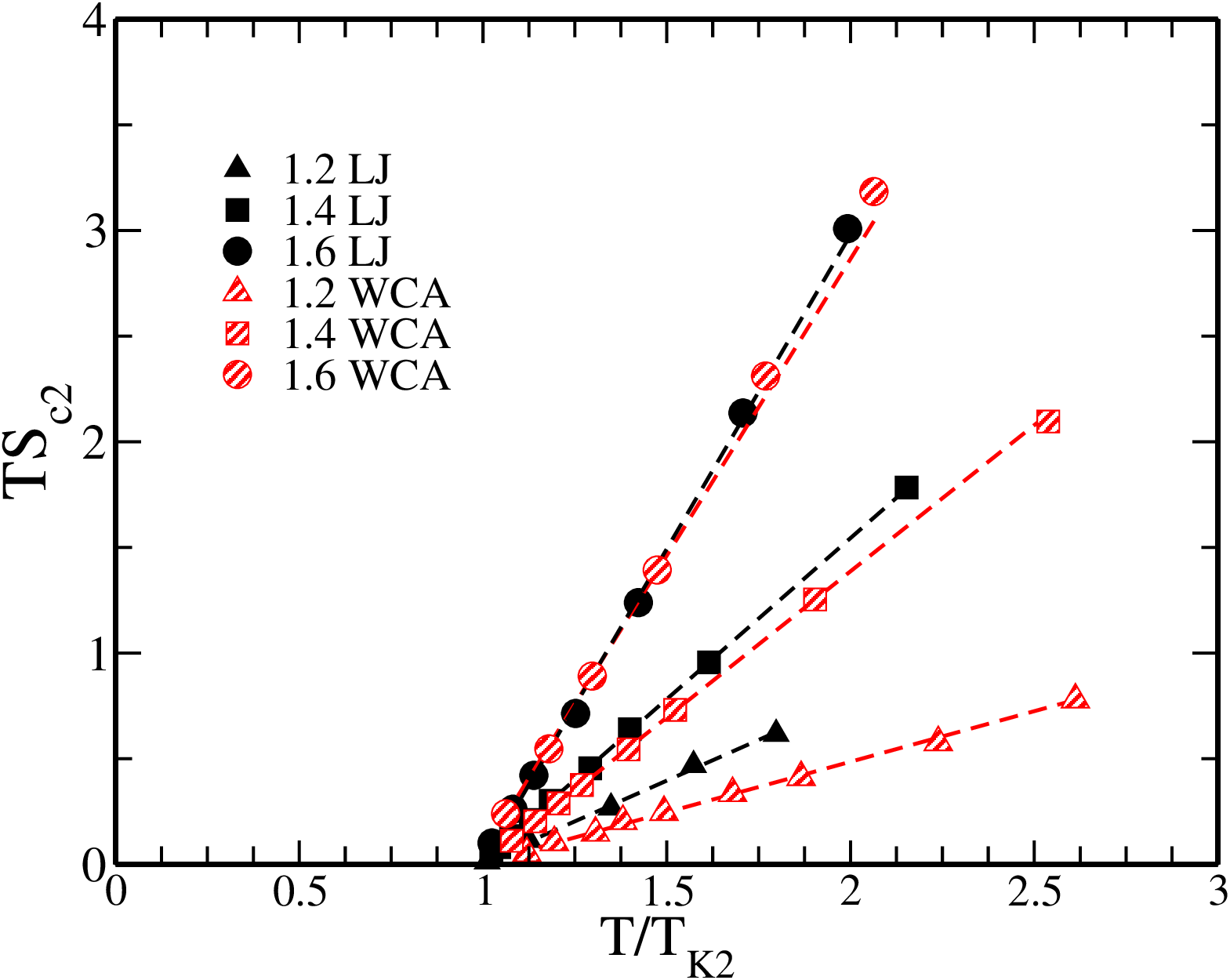}
\caption{ $TS_{c2}$ vs. $T/T_{K2}$ plot. Determination of pair thermodynamic fragility, $K_{T2}$ from the slope of the linear fit for all the systems. The temperature ($T_{K2}$) where $S_{C2}$ goes to zero, 
is tabulated in Table \ref{pair-thermodynamic-fragilities-table}.}
\label{pair-thermodynamic-fragilities-plot}
\end{figure}

\begin{table}
\caption{The values of pair thermodynamic fragility and  pair Kauzmann temperature  for all three densities are tabulated below. 
At higher density both the values of $K_{T2}$ and $T_{K2}$  
come closer for LJ and WCA 
systems.}
 \begin{tabular}{|c|c|c|}
    \hline
    $\rho$ & $K_{T2} (LJ)$  &$K_{T2} (WCA)$  \\ \hline
    1.2 &0.795&0.483\\ \hline
    1.4 &1.555 & 1.358  \\ \hline
    1.6 &2.971& 2.936\\ \hline
\end{tabular}
 \begin{tabular}{|c|c|c|}
    \hline
    $\rho$ & $T_{K2} (LJ)$  &$T_{K2} (WCA)$  \\ \hline
    1.2 &0.445&0.268\\ \hline
    1.4 &0.929 & 0.788  \\ \hline
    1.6 &1.76& 1.69\\ \hline
\end{tabular}
\label{pair-thermodynamic-fragilities-table}
\end{table}

\subsection{Dynamics predicted by pair correlation}
\begin{figure}[h]
\centering
\includegraphics[width=0.5\textwidth]{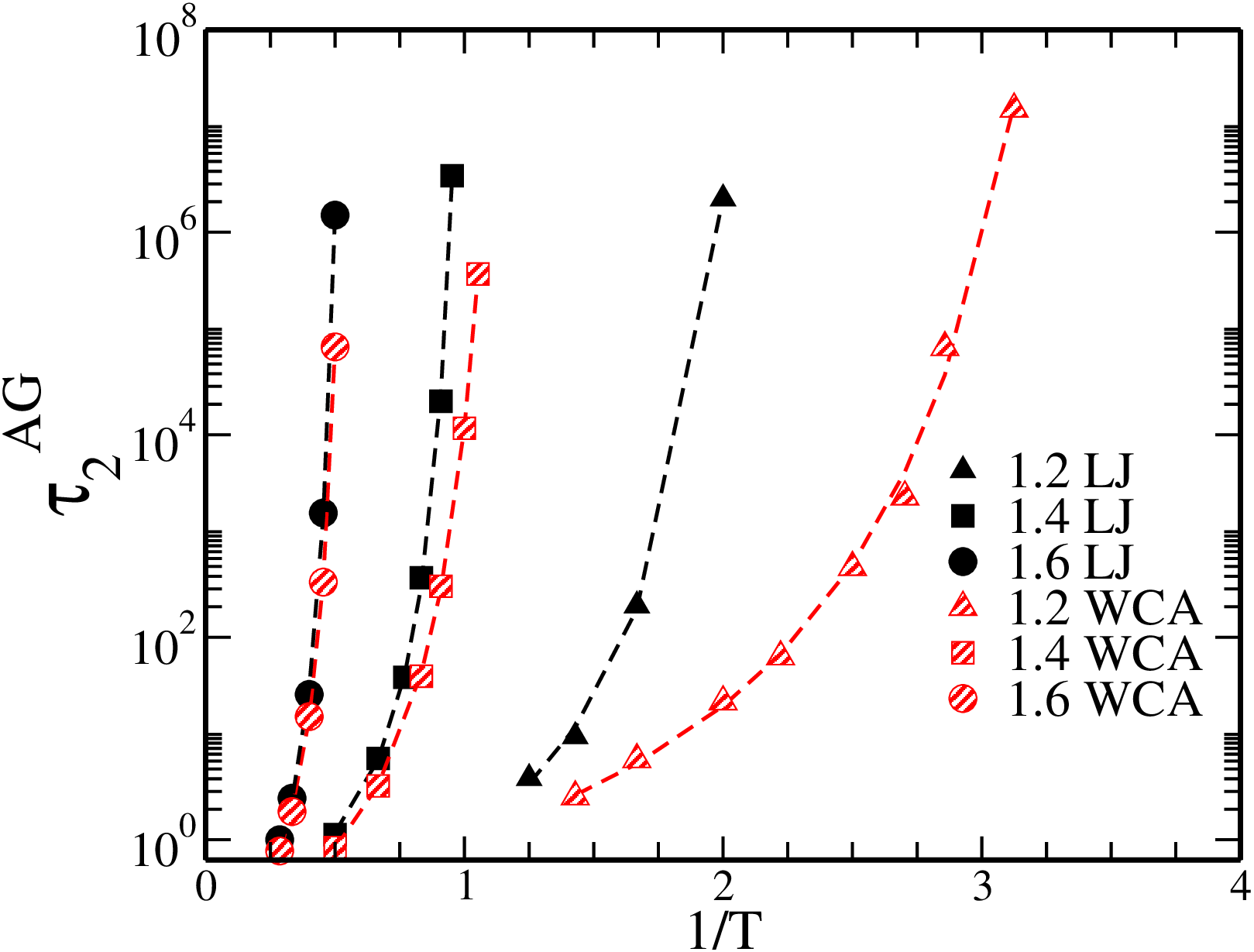}
\caption{ The temperature dependence of$\tau_2^{AG}$ for all the systems. The difference in the dynamics for LJ and WCA systems is driven by two-body correlation.
 Similar to the actual relaxation time, the plot shows the difference decreases as density increases. }
\label{tau2_ag_plot}
\end{figure}

Once we have determined the role of pair correlation in entropy, we now
 determine its effect on the low temperature dynamics. The AG relation can be re expressed as \cite{bssb}, 
\begin{eqnarray}
\tau(T)&=&\tau_{o}\exp\left(\frac{A}{TS_{c}}\right)=\tau_{o}\exp\left(\frac {A}{TS_{C2}}-\frac{A\Delta S }{TS_{C2}S_{C}}\right) \nonumber\\
&=&\tau^{AG}_{2} \exp\left(-\frac{A\Delta S }{TS_{C2}S_{C}}\right), 
\label{equ_tau_estm}
\end{eqnarray}
\noindent
where $\tau^{AG}_{2}(T) =\tau_{o}\exp\left(\frac {A}{TS_{C2}}\right)$. 
Note that the equalities in Eq.\ref{equ_tau_estm} are exact as long as $T, S_{C2}, S_{C}$ are finite.
The $\tau^{AG}_{2}$ for the LJ and the WCA systems are plotted in Fig.\ref{tau2_ag_plot} 
for all the densities.
The density and system dependence of $\tau^{AG}_2$ also shows similar characteristics as the simulated relaxation time, $\tau$. At lower 
density $\tau^{AG}_2$ of the two systems are far apart and as $\rho$ increases, they come closer.
 We also find that for a fixed density, $\tau^{AG}_{2}$ diverges at higher temperature compared to the actual relaxation time. Although this behaviour is 
 reminiscent of  MCT prediction \cite{tarjus_pre}, in the following part of the discussion we will show that this theory works in a regime where 
 the MCT and other microscopic theories which rely on the
 pair-correlation ($g(r)$) fail to capture the difference between the dynamics of the two systems\cite{tarjus_pre,Schweizer1,Schweizer2}.
  The pair kinetic fragility , $K_{VFT2}$, as obtained by fitting the temperature dependence of  $\tau^{AG}_{2}$  to a VFT form,
 shows that the LJ system is more fragile.
If we compare the pair fragility to the total fragility of the system we find that both for the kinetic and thermodynamic fragilities the pair one predicts a higher 
value. Usually the fragility of a system is connected to many body interaction \cite{tarjus-fragility-arXiv}.
However our analysis shows a connection of fragility to the two body correlations which is similar to a recent observation 
by Wang {\it et al} \cite{Wang-soft-matter}.

In an earlier work by some of us performed at a fixed density, we have shown that in contradiction to conventional wisdom the divergence in dynamics and also  
 the difference between the dynamics in the LJ and WCA systems are driven by pair correlation. Our present study over a wider range of density confirms
this as a generic behaviour independent of density of the systems.

\begin{figure}[h]
\centering

\includegraphics[width=0.49\textwidth]{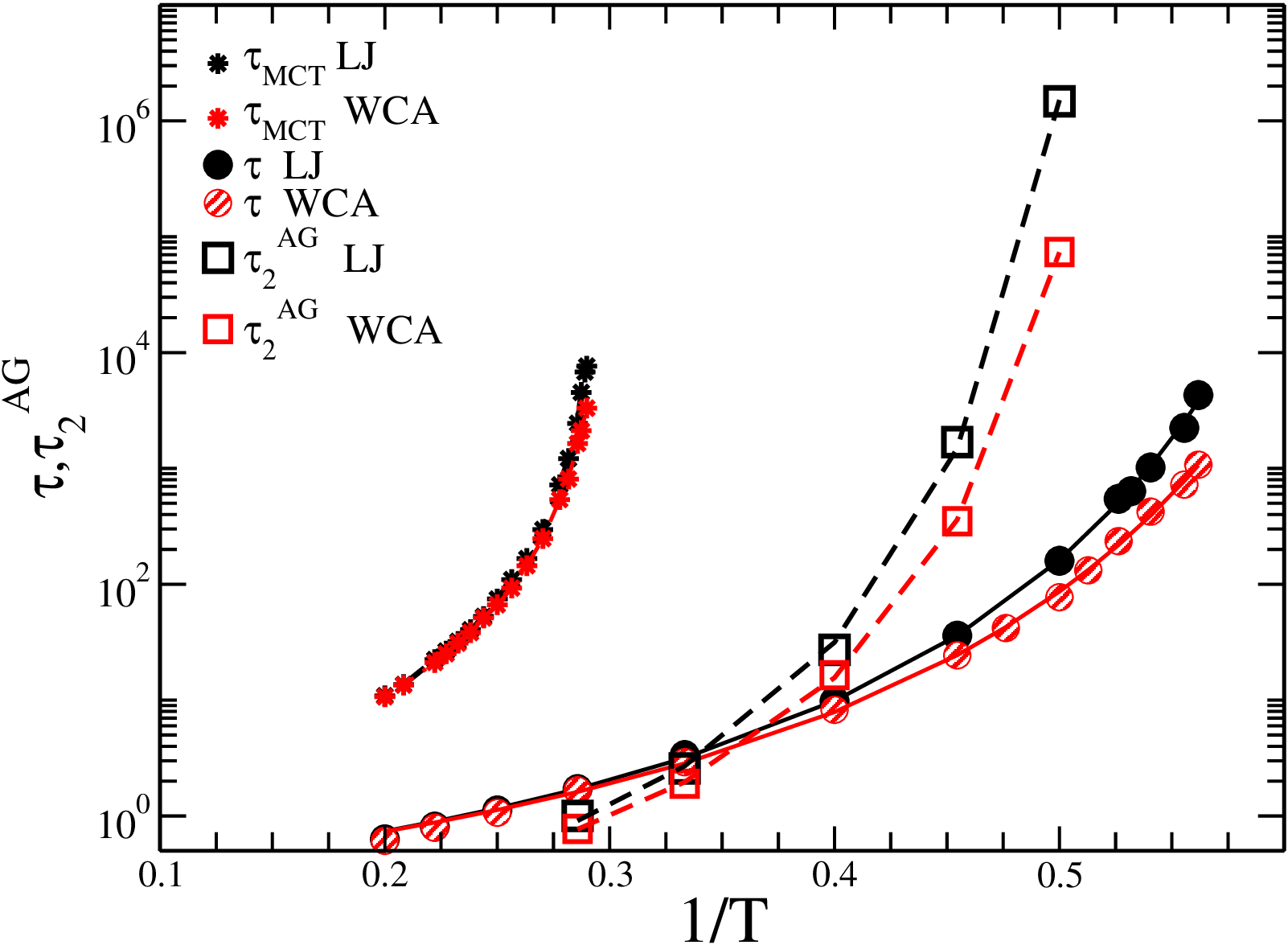}
\caption{ $\tau$, $\tau_{2}^{AG}$,$\tau_{MCT}$ vs. $1/T$ for LJ and WCA systems at $\rho=1.6$. Unlike MCT relaxation time, the estimated $\tau_{2}^{AG}$ shows a difference in the dynamics of 
LJ and WCA systems.
}
\label{tau_tau2_ag_1.6}
\end{figure}

  We next focus on  $\rho=1.6$ where the $\tau_{MCT}$ (calculation details given in Sec. IIIF) for both the systems almost overlap. In Fig.\ref{tau_tau2_ag_1.6} 
we show that even at
this density $\tau_2^{AG}$ values are different and the difference is more than that found for the total relaxation time.
A probable reason for the AG theory, only with pair correlation, $S_2$, to work better
 than
 the MCT is that the AG relation uses the entropy.
 In the calculation of the two body entropy $S_{2}$ although we consider only the radial distribution function but in effect it 
calculates a many body quantity. This in turn magnifies the effect observed at the two body level. 
      
     However  although $\tau^{AG}_{2}$ as compared to $\tau_{MCT}$ is shown to be better in predicting the difference between the two systems, it shows a divergence at a higher temperature reminiscent of the behaviour of $\tau_{MCT}$. 
Surprisingly the two behaviours are similar in trend but they are fundamentally different, where the AG comes from an activation picture whereas MCT in the form
presented here is a mean field theory at the two body level without any activation.
  In a recent work by us it was shown that those two fundamentally different theories predict the correct temperature dependence of relaxation
time in a certain temperature region \cite{unravel}. However it has also been shown that the AG theory at the two body level cannot predict the MCT power law behaviour .
Even above $T_c$, the residual multiparticle entropy plays an important role and below $T_c$, RMPE determines the dymanics fully. These findings provide
 the evidence that multiparticle correlations are
essential to correctly describe the physics of supercooled liquids. The effect of density on multiparticle correlation will be addressed in future work.

   \begin{figure}[h]
\centering
{
\includegraphics[width=0.5\textwidth]{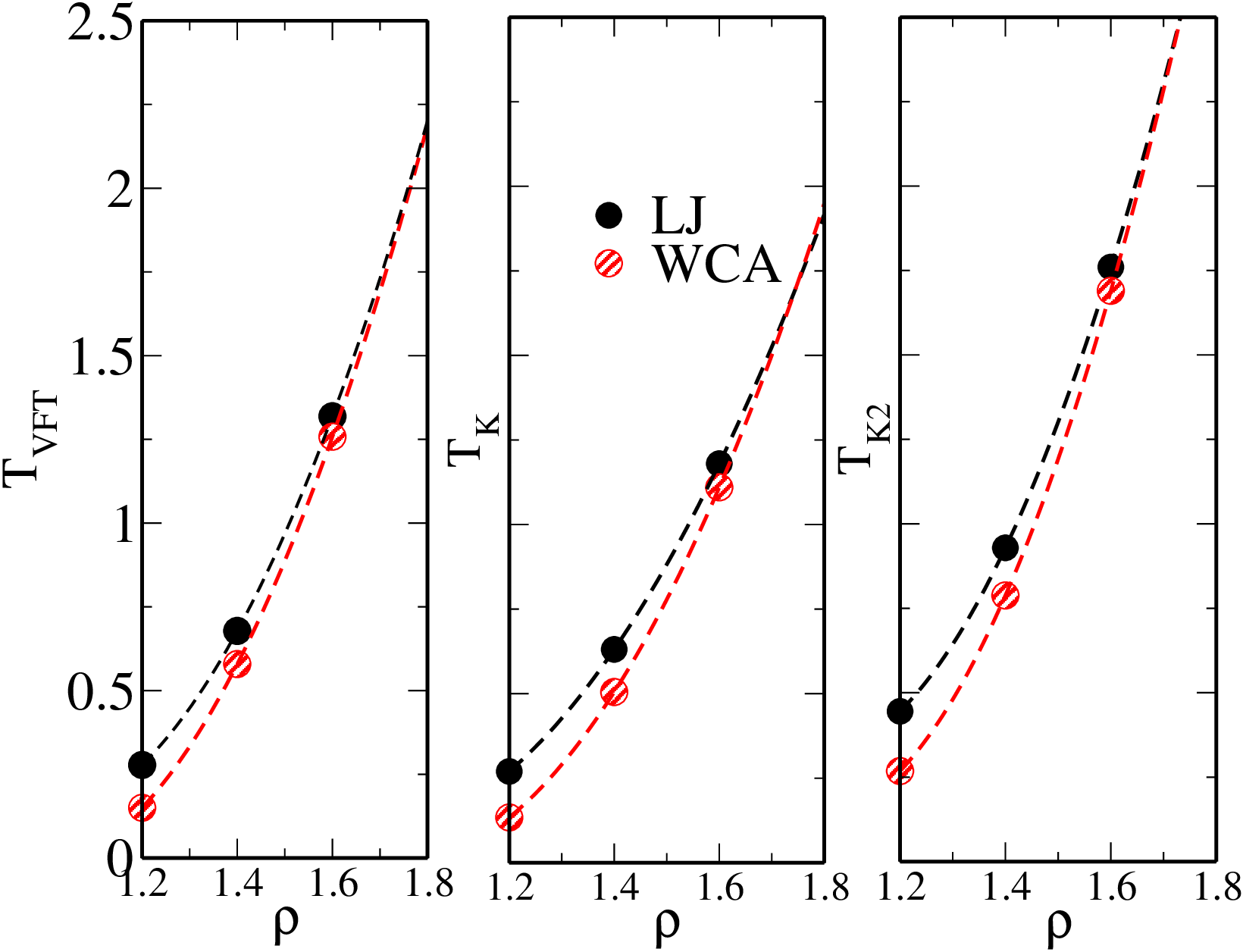}}
\caption{ Density dependence of $T_{VFT}$, $T_K$, $T_{K2}$ have been fitted and extrapolated to  a higher density, which show all three temperatures
predicts that LJ and WCA systems will behave in a similar way around $\rho=1.8$.   }
\label{divergence_temp}
\end{figure} 

\subsection{Phase diagram from thermodynamics and kinetics}

In this section by analyzing the density dependence of the different
divergence temperatures calculated in the previous sections, we make an attempt to predict the density where the thermodynamical and dynamical properties of 
both the LJ and WCA systems will be similar.
 The density dependence of the divergent temperatures behave in a similar fashion. 
 As mentioned earlier $T_{VFT}$, $T_{K}$, $T_{K2}$ values come closer as density increases (Fig.\ref{divergence_temp}). 
If we fit the data and extrapolate it to higher densities, we find that all the three temperatures predict that the WCA and LJ system will behave in a similar way 
around $\rho\simeq1.8$.  Wang {\it et. al} have shown that above $\rho=1.9\pm 0.1$ the  structural and vibrational quantities of the glasses show scaling collapse \cite{Wang-Xu-prl}.
Another study by Wang {\it et. al} have shown that at high density ($\rho \simeq 1.8$) both the LJ and WCA systems have similar structure, fragility and dynamics\cite{Wang-soft-matter}. 
The MCT critical temperature, which has been earlier connected to the pair Kauzman temperature \cite{unravel},
also shows that  at higher densities the difference decreases\cite{tarjus_prl} .
Dyre and coworkers have claimed that  the inclusion of interactions of all first coordination shell neighbours \cite{Toxvaerd}, can make LJ and WCA systems similar.
This however predicts a density which is higher than that predicted by the VDOS or the divergent temperatures.
Note that the density predicted by all the different divergence temperatures are similar.
Our finding, that above a certain density the dynamics and thermodynamics of both LJ and WCA systems will behave 
in a similar manner, is consistent with the observation made earlier \cite{Wang-Xu-prl,Wang-soft-matter,tarjus_prl,Toxvaerd}.

\section{Conclusion}
In this work, we study a wide range of  densities and show that 
 the difference in dynamics for the LJ and thr WCA systems is a manifestation of their thermodynamical difference.
 As density increases
the differences in the dynamical and thermodynamical quantities of the LJ and the WCA systems decrease, 
as a result of which the fragilities and the transition temperatures come closer at higher densities.
The well known Adam Gibbs relation, which connects the thermodynamic and kinetic fragilities, is found to be valid for all the densities
and systems studied here. 
 Our study shows that although the total pair correlation function for the LJ and the WCA systems are similar, the pair configurational entropy and corresponding dynamics show
a large difference.
 Our analysis from partial radial distribution function and partial pair entropy show that despite the maximum difference in BB rdf, due to the lower mole fraction of B type of
 particles, the difference in 
pair entropy comes from AB and AA pairs . We also find that the difference in the pair 
entropy arises from the short range part of the correlation functions.
The density dependence of the pair configurational entropy and the two body relaxation time follows the same trend as observed for the corresponding total quantities. 

In an earlier work done at one particular density we have shown that the difference in thermodynamics between the LJ and the WCA systems and also the divergence of their
dynamics are driven by pair correlation\cite{bssb}. In this study, we show that this indeed is a generic behaviour independent of the density of the system.
 We find that  both the thermodynamic and kinetic
fragility predicted by pair correlation are
larger than that predicted by the total correlation. Usually the fragility of a system is thought to be connected to cooperativity  and thus to a collective behaviour \cite{coslovich}.
However our analysis shows that the fragility is also connected to the two body correlation which is similar to a recent observation 
by Wang {\it et al} \cite{Wang-soft-matter}.

The dynamics predicted by the AG relation at the two body level is reminiscent of the MCT behaviour predicting a divergence of the dynamics at a 
higher temperature. However our analysis at $\rho=1.6$ shows that unlike MCT the pair correlation entropy and the corresponding dynamics can predict the difference
between the two systems. Thus AG relation at the two body level, although has the shortcomings of the MCT, fairs better than MCT in predicting the difference
between the two systems. A probable reason for this may be the fact that entropy even when calculated at the two body level provides a many body quantity.

The analysis of the density dependence of the different divergent temperatures predict that around $\rho\simeq1.8$ both systems will behave in a similar manner.
This supports the earlier observations \cite{Wang-Xu-prl,Wang-soft-matter,tarjus_prl,Toxvaerd}.
Our study although emphasizes the role of pair correlation, we also find that to describe the total dynamics pair correction is not enough and higher
order terms are important.

 \section{Acknowledgements}
This work has been supported by the Department of Science and Technology (DST), India and CSIR-Multi-Scale Simulation and Modeling project.
 AB thanks CSIR and MKN thanks UGC for fellowship. SS thanks the Sheikh Saqr Laboratory, JNCASR, for support.


\end{document}